\documentclass[]{lmcs}
\usepackage[utf8]{inputenc}
\pdfoutput=1

\usepackage{lastpage}
\lmcsdoi{19}{1}{5}
\lmcsheading{}{\pageref{LastPage}}{}{}%
{Dec.~01,~2020}{Jan.~18,~2023}{}

\newif\ifproofs%
\proofstrue%

\usepackage{wrapfig}

\usepackage{color}
\definecolor{boxshade}{gray}{0.85}
\usepackage{latexsym}
\usepackage{ifthen}
\usepackage{verbatim}

\title[Survey on Param. Verification with Threshold Automata and ByMC]{Survey on Parameterized Verification with Threshold Automata
  and the Byzantine Model Checker}

\author[I.~Konnov]{Igor Konnov\lmcsorcid{0000-0001-6629-3377}}[a]
\address{Informal Systems, Vienna, Austria}
\email{igor@informal.systems, ilina@informal.systems, josef@informal.systems}
\author[M.~Lazi\'c]{Marijana Lazi\'c\lmcsorcid{0000-0002-9222-6191}}[b]
\address{TU Munich, Munich, Germany}
\email{lazic@in.tum.de}
\author[I.~Stoilkovska]{Ilina Stoilkovska}[a,c]
\address{TU Wien, Vienna, Austria}
\email{stoilkov@forsyte.at}
\author[J.~Widder]{Josef Widder\lmcsorcid{0000-0003-2795-611X}}[a]
\thanks{%
    M.~Lazi\'c was supported  by the European Research Council (ERC) under the European Union's Horizon 2020 research and innovation programme under grant agreement No~787367 (PaVeS).
    I.~Konnov, I.~Stoilkovska and J.~Widder were supported by Interchain Foundation (Switzerland).
    I.~Stoilkovska was also supported by the Austrian Science Fund (FWF) via the Doctoral College LogiCS W1255.
}

\usepackage{enumitem}
\setlist{nolistsep}

\usepackage{multicol}

\usepackage{mathtools}
\usepackage{amsmath}
\usepackage{hyperref}
\usepackage{listings}
\usepackage{environ}
\usepackage{csvsimple}
\usepackage{longtable} \usepackage[normalem]{ulem} 
\usepackage{fourier-orns} \usepackage{wasysym} \usepackage{marvosym}        \usepackage{bbding} \usepackage{pifont}

\usepackage{units}

\usepackage{amsfonts,amssymb,amsmath}

\usepackage{tikz}
\usetikzlibrary{arrows}
\usetikzlibrary{shapes}
\usetikzlibrary{calc}
\usetikzlibrary{positioning}
\usetikzlibrary{patterns}
\usetikzlibrary{angles}

\usepackage{array}
\newcounter{rowno}
\usepackage{amssymb,amsmath}

\usepackage{listings}
\lstdefinelanguage{promela}
  {morekeywords={do,od,init,proctype,for,if,fi,else,goto,byte,int,bool,bit,chan,mtype,atomic,d_step,nempty,empty,break,skip,active,ltl,symbolic,assume,some,all,card,havoc},
  morecomment=[l]{//}, morecomment=[s]{/*}{*/},mathescape=true,escapechar={@},
  basicstyle=\sffamily\small, commentstyle=\itshape\rmfamily\small\color{black!50},
  keywordstyle=\sffamily\bfseries\small}

\lstdefinelanguage{pseudo}
  {morekeywords={init,with,or,if,then,else,fi,and,not,while,do,od,done,
    distinct,
    case, goto,local,algorithm, function, for, each, times, from, to,
    variables, procedure, recursive, return, send, all, wait, until, input,
    decide, received, report, break, gather, receive},
  morecomment=[l]{//}, morecomment=[s]{/*}{*/},
  mathescape=true,escapechar={@},
  basicstyle=\sffamily,commentstyle=\itshape\rmfamily,keywordstyle=\sffamily\bfseries,numbers=left
}
\lstdefinelanguage{threshauto}
  {morekeywords={thresholdAutomaton,local,shared,parameters,assumptions,locations,inits,
    rules,when,do,unchanged,specifications,true},
  morecomment=[l]{//}, morecomment=[s]{/*}{*/},mathescape=true,escapechar={@},
  basicstyle=\sffamily\small, commentstyle=\itshape\rmfamily\small\color{black!50},
  keywordstyle=\sffamily\bfseries\small}

\usepackage{pifont}
\newcommand{\DD}{\ding{54}}
\newcommand{\CRO}{\textsc{c0}}
\newcommand{\CRI}{\textsc{c1}}

\newcommand{\VO}{\textsc{v0}}
\newcommand{\VI}{\textsc{v1}}

\renewcommand{\AC}{\textsc{ac}}
\newcommand{\SE}{\textsc{se}}
\newcommand{\IO}{\textsc{i0}}
\newcommand{\JO}{\textsc{j0}}
\newcommand{\II}{\textsc{i1}}
\newcommand{\JI}{\textsc{j1}}
\newcommand{\SR}{\textsc{sr}}
\newcommand{\SP}{\textsc{sp}}
\newcommand{\EO}{\textsc{e0}}

\newcommand{\EI}{\textsc{e1}}
\newcommand{\DI}{\textsc{d1}}
\newcommand{\CTO}{\textsc{ct0}}
\newcommand{\CTI}{\textsc{ct1}}

\gdef\dash---{\thinspace---\hskip.16667em\relax}
\gdef\ndash---{\thinspace--\hskip.16667em\relax}

\newcommand{\ruleset}{{\mathcal R}}

      \newcommand{\local}{{\mathcal L}}

\newcommand{\PrecondU}{\Phi^{\mathrm{rise}}}
\newcommand{\PrecondL}{\Phi^{\mathrm{fall}}}

\newcommand{\Ctx}{\Omega}

\newcommand{\rulesliceclass}{\raise0ex\hbox{$(\slice{\ruleset}{\Ctx})$}\big/\lower.5ex\hbox{\hskip-.15em{{\scriptsize $\sim$}}}}

\newcommand{\true}{\textit{true}}

\newcommand{\newreftheorem}[2]{\newenvironment{#1}[1]{\par\vspace{3mm}\noindent\textbf{#2~\ref{##1}.}
\em}{\rm}
}
\newreftheorem{reflemma}{Lemma}
\newreftheorem{refproposition}{Proposition}
\newreftheorem{reftheorem}{Theorem}
\newreftheorem{refcorollary}{Corollary}

\newcommand{\cpp}[1]{#1\texttt{\footnotesize{++}}}

\newcommand{\counters}{{\vec{\boldsymbol\kappa}}}

\newcommand{\var}[1]{\textsf{#1}}
\newcommand{\guard}{\varphi}

\newcommand\ltlF{\textsf{\textbf{F}}\,}
\newcommand\ltlG{\textsf{\textbf{G}}\,}

\newcommand\LTLX{$\mbox{\textsf{LTL}}_{\textsf{{-X}}}$}

\newcommand\ELTLTB{$\mbox{\textsf{ELTL}}_{\textsf{FT}}$}

\renewcommand\vec[1]{\mathbf{#1}}

\newcommand\tbh[1]{\textsf{\textbf{\scriptsize{#1}}}}

\newcommand\slice[2]{#1|_{#2}}

\usepackage{color}

\usepackage{ifthen}

\usepackage{colortbl}
\definecolor{highlight}{gray}{0.85}
\newcolumntype{h}{>{\columncolor{highlight}}c}
\definecolor{crcolor}{HTML}{FFA07A}

\newcommand{\arule}{r}

\lstdefinelanguage{distal}
  {morekeywords={case,class,extends,val,var,UPON,RECEIVING,START,WITH,
      TIMES,DO,IF,THEN,SEND,TO,ALL,False,True},
  morecomment=[s]{/*}{*/}, morecomment=[l]{//},
  mathescape=true,escapechar={@},
  basicstyle=\sffamily\small,
  commentstyle=\itshape\rmfamily\small, numbers=left, numberstyle=\tiny,
  xleftmargin=2em, framexleftmargin=1.5em
}

\newcommand{\floodset}{\textsf{FloodSet}}
\newcommand{\faircons}{\textsf{FairCons}}
\newcommand{\phaseking}{\textsf{PhaseKing}}
\newcommand{\phasequeen}{\textsf{PhaseQueen}}
\newcommand{\hybridking}{\textsf{HybridKing}}
\newcommand{\hybridqueen}{\textsf{HybridQueen}}
\newcommand{\byzking}{\textsf{ByzKing}}
\newcommand{\byzqueen}{\textsf{ByzQueen}}
\newcommand{\omitking}{\textsf{OmitKing}}
\newcommand{\omitqueen}{\textsf{OmitQueen}}
\newcommand{\floodmin}{\textsf{FloodMin}}
\newcommand{\floodminomit}{\textsf{FloodMinOmit}}
\newcommand{\ksetomit}{\textsf{kSetOmit}}

\newcommand{\hybridrb}{\textsf{HybridRB}}
\newcommand{\omitrb}{\textsf{OmitRB}}

\newcommand{\strb}{\textsf{STRB}}

\newcommand{\benor}{\textsf{Ben-Or}}

\newcommand{\tlap}[0]{\textsc{TLA}\textsuperscript{+}}

\newcommand{\ropen}[1]{[#1)} 

\begin{document}

\begin{abstract}
Threshold guards are a basic primitive of
     many fault-tolerant algorithms that solve classical problems in
     distributed computing, such as reliable broadcast, two-phase
     commit, and consensus.
Moreover, threshold guards can be found in recent blockchain
     algorithms such as, e.g., Tendermint consensus.
In this article, we give an overview of
techniques for automated verification of
threshold-guarded fault-tolerant distributed algorithms,
implemented in the Byzantine Model Checker
(ByMC).
These threshold-guarded algorithms have the following
features:
(1)~up to~$t$ of processes may crash or behave Byzantine;
(2)~the correct processes count messages and make progress
when they receive sufficiently many messages, e.g.,
at least $t+1$;
(3)~the number~$n$ of processes in the system is a parameter,
as well as the number~$t$ of faults; and
(4)~the parameters are restricted by a resilience condition,
e.g., $n > 3t$.
Traditionally, these algorithms were implemented in distributed
     systems with up to ten participating processes.
Nowadays, they are implemented in distributed systems that  involve
     hundreds or thousands of processes.
To make sure that these algorithms are still correct for that scale,
     it is imperative to verify them for all  possible values of the
     parameters.
\end{abstract}

\maketitle

\section{Introduction}

\paragraph{Distributed Systems.}

The recent advent of blockchain technologies~\cite{Nak08,DeckerSW16,AbrahamMNRS16,Buchman2016,YinMRGA19,But2014}
has brought fault-tolerant distributed
     algorithms into the spotlight of computer science and software
     engineering.
Consider a blockchain system, where
a blockchain algorithm ensures coordination
of the participants (i.e., the processes) in the system.
We observe that to achieve coordination, the processes need to solve a
     coordination problem called \emph{atomic
     (or, total order) broadcast}~\cite{Hadzilacos:1993},
that is, every
     process has to execute the same transactions in the same order.
To achieve that, the algorithm typically relies on a
\emph{resilience condition} that
     restricts the fraction of processes that may be
     faulty~\cite{PeaseSL80}.
These classic Byzantine fault tolerance concepts are indeed
at the base of several modern blockchain systems, such as
Tendermint~\cite{Buchman2016} and HotStuff~\cite{YinMRGA19}.

In addition to the above mentioned practical importance of distributed systems,
the reasons for the long-standing interest~\cite{LeLann77,Lamport78,PeaseSL80,FLP85} in
      this field is that distributed consensus is non-trivial in
      two~aspects:
\begin{enumerate}

\item\label{hard:2}
     Most coordination problems are
     impossible to solve without imposing constraints on the
     environment, e.g., an upper bound on the fraction of faulty
     processes, assumptions on the behavior of faulty processes, or
     bounds on message delay and processing speed (i.e.,
     restricting interleavings)~\cite{PeaseSL80,FLP85,Dolev:1987}.

\item Designing correct solutions is hard, owing to the
     huge state and execution space, and
     the complex interplay of the assumptions about the faulty processes,
     message delivery, and other environment assumptions mentioned above.
 Therefore, it is not surprising that even published protocols
       may contain bugs, as reported,
     e.g., by~\cite{LR93,MS06}.
\end{enumerate}

\noindent
Due to the well-known impossibility of asynchronous fault-tolerant
     consensus (FLP impossibility)~\cite{FLP85},
     much of the distributed systems research focuses on:
     (a)~what kinds of problems are solvable in asynchronous systems (e.g.,
     some forms of reliable broadcast) or
     (b)~what kinds of systems allow to solve consensus.

     \paragraph{Computer-aided Verification.}

Due to the huge amount of funds managed by blockchains, it is
     crucial that their software is free of bugs.
At the same time, these systems are characterized by a large number of
     participants, which makes the application of
automated verification methods inevitably run into the well-known
     state space explosion problem.
This is why, in such a system, consisting of an a priori unknown
number of participants, it is desirable to show the system correctness
using parameterized verification techniques.
However, the well-known undecidability results for the
     verification of parameterized
     systems~\cite{AK86,S88,EN95,E97,2015Bloem} also apply in this setting.
One way to circumvent both the state space explosion problem and
the undecidability of parameterized verification is to develop domain specific
     methods that work for a specific subclass of systems.

In this article, we survey verification techniques for fault-tolerant
     distributed algorithms, which:
     (a)~for asynchronous systems deal with the concepts of
     broadcast and atomic broadcast under resilience conditions,
     and
     (b)~can be used to verify consensus in synchronous and randomized
     systems.
While the benchmarks we used to evaluate these verification techniques
predominantly come from the classic fault-tolerant distributed systems
literature,
we note that they in part address the challenges imposed by distributed
systems existing in the real world
(such as the two blockchain systems mentioned above).
We present several verification methods that have been developed in
the threshold-automata framework~\cite{KVW17:IandC}.
This framework has been
introduced as a succinct formalization of the transition relations of
threshold-based fault-tolerant distributed algorithms.

The remainder of the paper is structured as follows.
In Sections~\ref{sec:techniques} to~\ref{sec:multi},
we survey some of the most
     fundamental system assumptions that allow to solve problems in
     distributed computing in the presence of faults, and introduce example algorithms.
We consider different computational
     models; namely, synchronous lock-step systems, asynchronous
     systems, and probabilistic systems.
We also discuss
     how these algorithms (operating in either of the three
     computation models) can be formalized using threshold automata and
     how they can be automatically verified.
Since threshold automata represent an abstraction of distributed
     algorithms,
in Section~\ref{sec:modeling} we discuss how this abstraction can be
     automatically generated from a formalization close to the
     algorithm descriptions in the literature.
Then, in Section~\ref{sec:bymc}, we present how our tool ByMC evolved in
     the last years and which techniques were implemented.
In Section~\ref{sec:Tendermint} we demonstrate  how ByMC can be used
     to analyze Tendermint, a state-of-the art consensus algorithms
     used in the Cosmos blockchain ecosystem.
And finally, in Section~\ref{sec:related}, we give an overview of
     related work.

\section{Parameterized Verification of Synchronous Algorithms}%
\label{sec:techniques}%
\label{sec:syncTA}

\subsection{Synchronous Algorithms}%
\label{sec:syncalg}
In a synchronous algorithm,
all correct processes execute the algorithm code line-by-line in lock-step.
The computation is organized in rounds, which consist of three steps:
send-to-all, receive, and local computation.
A message sent by a correct process to a correct process is received
     within the same round.
After sending and receiving messages in lock-step, all correct
     processes continue by evaluating the guards, before they all
     proceed to the next round.
Because this semantics ensures that all processes move together, and all
     messages sent by correct processes
     are received within the same round, no
     additional fairness constraints are needed to ensure liveness
     (i.e., that something good eventually happens).
In practice, this approach is often considered slow and expensive,
     as it has to be implemented with timeouts that are
     aligned to worst case message delays (which can be very
     high in real networks).
However, the synchronous semantics offers a high-level
     abstraction that allows one to design algorithms easier.

We now present an example synchronous algorithm from the literature,
which we refer to as \floodmin~\cite{ChaudhuriHLT00}.
Its pseudocode is given in Figure~\ref{fig:FloodMin} on the left.
This algorithm is run by~$n$ replicated processes, up to~$t$ of which
     may fail by crashing, that is, by prematurely halting.
It solves the $k$-set agreement problem, that is, out of the~$n$
     initial values, the goal of each process
     is to decide on one value, such that the
     number of different decision values is at most~$k$.
By setting~$k=1$, we obtain that there can be exactly one decision
     value, which coincides with the definition of consensus.
For simplicity of presentation, we consider
the case where $k=1$ and where the set of initial values is the
set $\{0, 1\}$.

Each process running \floodmin\ has a variable
$\var{best}$, which stores its input value, coming from the set $\{0, 1\}$
(line~\ref{floodmin:input}).
One iteration of the loop at line~\ref{floodmin:forloop}
is called a \emph{round}.
In each round, each process broadcasts its value $\var{best}$
in line~\ref{floodmin:broadcast}.
The variable $\var{best}$ is updated in each round as the
minimum of all received values (line~\ref{floodmin:update}).
\floodmin\ runs for $\lfloor t/k \rfloor + 1$ rounds (line~\ref{floodmin:forloop}),
which in case of $k=1$ amounts to $t + 1$ rounds.
The number $\lfloor t/k \rfloor + 1$ of rounds ensures
that there is at least one clean round in which at most $k-1$
processes crash.
When we consider consensus (i.e., when $k=1$), this means there is a round in which no
process crashes, such that all processes receive the same values
$b_1, \dots b_\ell$.
As a result, during that round, all processes set
$\var{best}$ to the same~value.
Finally, after the loop line~\ref{floodmin:forloop} terminates,
each process decides on the value of the variable $\var{best}$
(line~\ref{floodmin:decide}).

\subsection{Synchronous Threshold Automata}
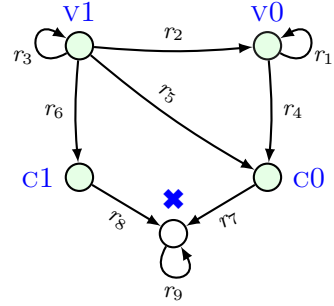
\begin{figure}[t]
\begin{minipage}{.49\textwidth}
\begin{lstlisting}[language=pseudo,columns=fullflexible]
best := input_value; @\label{floodmin:input}@
for each round 1 through @$\lfloor t/k \rfloor +1$@ do @\label{floodmin:forloop}@
  broadcast best;  @\label{floodmin:broadcast}@
  receive values @$b_1, \dots b_\ell$@ from others;  @\label{floodmin:receive}@
  best := min @$\{b_1, \dots b_\ell\}$@;  @\label{floodmin:update}@
done;
decide best;  @\label{floodmin:decide}@
\end{lstlisting}
     \end{minipage}
     \begin{minipage}{.5\textwidth}

\tikzstyle{node}=[circle,draw=black,thick,minimum size=3mm,font=\normalsize]
\tikzstyle{init}=[circle,draw=black!90,fill=green!10,thick,minimum size=3mm,font=\normalsize]
\tikzstyle{final}=[circle,draw=black!90,fill=red!10,thick,minimum size=3mm,font=\normalsize]
\tikzstyle{rule}=[->,thick]
\tikzstyle{post}=[->,thick]
\tikzstyle{pre}=[<-,thick]
\tikzstyle{cond}=[rounded corners,rectangle,minimum
  width=1cm,draw=black,fill=white,font=\normalsize]
\tikzstyle{asign}=[rectangle,minimum
  width=1cm,draw=black,fill=gray!5,font=\normalsize]

\tikzset{every loop/.style={min distance=5mm,in=140,out=113,looseness=2}}
\begin{tikzpicture}[>=latex]

    \node[] at (1.25,  2.5)
     [init,label=above:\textcolor{blue}{$\VO$}] (V0) {};
    \node[] at (-1.25,  2.5)
     [init,node,label=above:\textcolor{blue}{$\VI$}] (V1) {};
    \node[] at (0, 0)
     [node,label=above:\textcolor{blue}{\DD}] (DD) {};

    \node[] at (1.25, .75)
     [init,node,label=right:\textcolor{blue}{$\CRO$}] (CR0) {};
    \node[] at (-1.25, .75)
     [init,node,label=left:\textcolor{blue}{$\CRI$}] (CR1) {};

\draw[post] (V1) edge[bend right=5] node[anchor=east]
    {\footnotesize{$r_6$}} (CR1);

\draw[post] (V1) edge[bend right=5] node[anchor=south, sloped, midway,
xshift={-2mm}]
    {\footnotesize{$r_5$}}(CR0);

\draw[post] (V1) edge[bend right=5] node[anchor=south, midway]
    {\footnotesize{$r_2$}}(V0);

\draw[post] (V0) edge[bend left=5] node[anchor=west]
    {\footnotesize{$r_4$}}(CR0);

\draw[post] (CR0) -- node[anchor=north, sloped, midway]
    {\footnotesize{$r_7$}}(DD);

\draw[post] (CR1) -- node[anchor=north, sloped, midway]
    {\footnotesize{$r_8$}}(DD);

\draw[rule] (V0) to[out=-30,in=30,looseness=8] node[anchor=west, near start]
    {\footnotesize{$r_1$}}(V0);

\draw[rule] (V1) to[out=210,in=150,looseness=8] node[anchor=east, near start]
    {\footnotesize{$r_3$}}(V1);

\draw[rule] (DD) to[out=-120,in=-60,looseness=8] node[anchor=north, midway]
    {\footnotesize{$r_9$}}(DD);

\end{tikzpicture}
      \end{minipage}
\caption{Pseudocode of \floodmin\ from~\cite{ChaudhuriHLT00} and its STA
     for $k=1$ and initial values from $\{0, 1\}$}%
     \label{fig:FloodMin}
\end{figure}

Synchronous threshold automata were introduced in~\cite{StoilkovskaKWZ19}.
A \emph{synchronous} \emph{threshold automaton}
\emph{(STA)}
that models the pseudocode of the algorithm \floodmin\ is given in
Figure~\ref{fig:FloodMin} on the right.
It resembles a control flow graph and is used
to encode the body of the loop on line~\ref{floodmin:forloop} in the
pseudocode.
The nodes of the STA are called \emph{locations}, and encode the
value of the process local variables and the program counter.
The directed edges of the STA are called \emph{rules}.
A rule models one iteration of the loop, such that
the two locations it connects reflect
the values of the local variables of a process
before and after the execution of one loop iteration.
The rules are labeled by \emph{threshold guards}, which encode
conditions from the pseudocode (and may be
parameterized by the number~$n$ of processes,
number~$f$ of faults, and the upper bound~$t$ on the number
of faults).

In the STA in Figure~\ref{fig:FloodMin},
the locations $\VO$ and  $\VI$ encode that a correct process has
the value $\var{best}$ set to 0 and 1, respectively.
The locations $\CRO$ and $\CRI$ analogously encode that a
crash-faulty process has the  value $\var{best}$ set to 0 and 1, respectively.
The location~\DD\ encodes that a crashed process stopped working and
does not restart.
We discuss how we model crash faults in more detail in Section~\ref{sec:crashfaults}.

By looking at the pseudocode,
we observe that a correct (resp.\ crash-faulty)
process sends a message containing the value~0,
if the value of its variable $\var{best}$ is~0
(line~\ref{floodmin:broadcast}).
In the STA, a correct (resp.\ crash-faulty)
process sends a message containing the value~0,
if it is in location~$\VO$ (resp.~$\CRO$).
Thus, to test how many processes have sent a message
containing the value~0, it is enough
to \emph{count} how many processes are in one of the locations
$\VO$ and~$\CRO$.
We denote by $\#\{\VO, \CRO\}$ the number of processes in
locations $\VO$ and~$\CRO$.

However, the pseudocode of~\floodmin\ contains conditions over the
number of \emph{received} messages (line~\ref{floodmin:receive}),
rather than the number of \emph{sent} messages.
As discussed before, in synchronous systems,
all messages sent by correct processes in a given round
are received by all correct processes in the same round.
On the contrary, not all messages sent by faulty processes
are received by all correct processes.
In the case of~\floodmin, as the crash-faulty processes may omit to
send messages to the correct processes, the number of received messages
may deviate from the number of correct processes that sent a message.
We will discuss this discrepancy in more detail in
Section~\ref{sec:modeling}.

To capture this in the STA, we proceed as follows.
Observe that in the pseudocode of~\floodmin,
the processes update their
value $\var{best}$ with the smallest received value
(line~\ref{floodmin:update}).
The correct processes that are in location~$\VO$ can
only stay in~$\VO$, which is captured by the rule~$\arule_1$.
The correct processes that are in location~$\VI$ can either:
\begin{enumerate}
    \item move to the location~$\VO$, if there is at least one message
containing the value~0, sent either by a correct or by a faulty process.
This is modeled by the rule~$\arule_2$,
whose guard~$\guard_1$ checks if $\#\{\VO, \CRO\} \ge 1$;
    \item stay in the location~$\VI$, if there is no message
    containing the value~0 sent by a correct process.
    This is modeled by the rule~$\arule_3$,
whose guard~$\guard_2$ checks if $\#\{\VO\} < 1$.
\end{enumerate}
Both the guards~$\guard_1$ and~$\guard_2$ can be satisfied in location~$\VI$.
This is used to model the non-determinism imposed by the crash-faulty
processes that manage to send messages only to a subset of the
other processes.
The rules $\arule_4, \arule_5$, and $\arule_6$ correspond to
the rules $\arule_1, \arule_2$, and $\arule_3$, respectively,
and are used to move processes that are crashing to
the locations where they are flagged as crash-faulty.
The rules $\arule_7$ and $\arule_8$ move the crash-faulty processes
to the location \DD, where they stay forever.

\begin{table}
\caption{The rules of the STA from Figure~\ref{fig:FloodMin}. We omit
the rules $r_1, r_4, r_{7}, r_{8}, r_9$ as they have the trivial guard
(\textit{true}).}%
\label{tab:floodminrules}
\small
\begin{center}
     \begin{tabular}{ll}
          \tbh{Rule} & \tbh{Guard} \\
               \hline
          $\arule_2$ & $\guard_1 \colon \#\{\VO, \CRO\} \ge 1$ \\
          $\arule_3$ & $\guard_2 \colon \#\{\VO\} < 1$\\
          $\arule_5$ & $\guard_1 \colon \#\{\VO, \CRO\} \ge 1$\\
          $\arule_6$ & $\guard_2 \colon \#\{\VO\} < 1$
     \end{tabular}
\end{center}
\end{table}

\subsection{Counter Systems}

The semantics of the STA is defined in
terms of a \emph{counter system}.
For each location in the STA, $\ell \in \{\VO, \VI, \CRO, \CRI, \mbox{\DD}\}$,
we have a counter~$\kappa[\ell]$ that stores the number of processes located in
location~$\ell$.
The counter system is parameterized in the
number~$n$ of processes,
number~$f$ of faults, and the upper bound~$t$ on the number
of faults.
Every transition in the counter system
updates the counters by moving all processes
simultaneously; potentially using a
different rule for each process,
provided that the guards of the applied rules
evaluate to true.

For the STA of the algorithm~\floodmin,
presented in Figure~\ref{fig:FloodMin},
in Figure~\ref{fig:fm-long} we depict an execution
in the counter system for parameter values
$n, t, f$ such that $n > t \ge f$ and $f > 3$.
Initially, there are $n-1$ processes in location $\VI$,
i.e., $\kappa_0[\VI] = n-1$, and
one process in location~$\CRO$, i.e., $\kappa_0[\CRO] = 1$.
Observe that these initial values of the counters make all the
rules in the STA enabled.
This is because all
the guards on all rules in the STA are satisfied by these
counters, as
$\#\{\VO, \CRO\} = \kappa_0[\VO] + \kappa_0[\CRO] = 1 \ge 1$ and
$\#\{\VO\} = \kappa_0[\VO] = 0 < 1$.
Thus, in any transition applied to~$\kappa_0$,
any rule that is outgoing of the locations $\VI, \CRO$
(which are the ones populated in~$\kappa_0$) can be taken by some
process.
In particular, the transition outgoing from $\kappa_0$ moves:
$n-2$ processes from $\VI$ to $\VI$ using the self-loop rule~$\arule_3$,
one process from $\VI$ to $\CRO$ using the rule~$\arule_5$,
  and
one process form $\CRO$ to \DD\ using the rule~$\arule_7$.

\subsection{Bounded Diameter}

\begin{figure}
     \scalebox{.8}{
\begin{tikzpicture}

    \node[] (conf0) at (.5, 5.25) {$\kappa_0$};
    \draw[step=1,black!80, very thin] (0, 0) grid (1, 5);

    \node[] (conf1) at (3.5, 5.25) {$\kappa_1$};
    \draw[step=1,black!80, very thin] (3, 0) grid (4, 5);

    \node[] (conf2) at (6.5, 5.25) {$\kappa_2$};
    \draw[step=1, black!80, very thin] (6, 0) grid (7, 5);

    \node[] (conff-1) at (9.5, 5.25) {$\kappa_{f-1}$};
    \draw[step=1, black!80, very thin] (9, 0) grid (10, 5);

    \node[] (conff) at (12.5, 5.25) {$\kappa_f$};
    \draw[step=1,black!80, very thin] (12, 0) grid (13, 5);

    \node[] (conff1) at (15.5, 5.25) {$\kappa_{f + 1}$};
    \draw[step=1,black!80, very thin] (15, 0) grid (16, 5);

    \node[blue] (v0) at (-.5, 4.5) {$\VO$};
    \node[blue] (v1) at (-.5, 3.5) {$\VI$};
    \node[blue] (c0) at (-.5, 2.5) {$\CRO$};
    \node[blue] (c1) at (-.5, 1.5) {$\CRI$};
    \node[blue] (f) at (-.5, .5) {\DD};

\node[] (c0v0) at (.5, 4.5) {0};
    \node[] (c0v1) at (.5, 3.5) {\small{$n-1$}};
    \node[] (c0c0) at (.5, 2.5) {1};
    \node[] (c0c1) at (.5, 1.5) {0};
    \node[] (c0f) at (.5, .5) {0};

\node[] (c1v0) at (3.5, 4.5) {0};
    \node[] (c1v1) at (3.5, 3.5) {\small{$n-2$}};
    \node[] (c1c0) at (3.5, 2.5) {1};
    \node[] (c1c1) at (3.5, 1.5) {0};
    \node[] (c1f) at (3.5, .5) {1};

\node[] (c2v0) at (6.5, 4.5) {0};
    \node[] (c2v1) at (6.5, 3.5) {\small{$n-3$}};
    \node[] (c2c0) at (6.5, 2.5) {1};
    \node[] (c2c1) at (6.5, 1.5) {0};
    \node[] (c2f) at (6.5, .5) {2};

\node[] (cf-1v0) at (9.5, 4.5) {0};
    \node[] (cf-1v1) at (9.5, 3.5) {\small{$n-f$}};
    \node[] (cf-1c0) at (9.5, 2.5) {1};
    \node[] (cf-1c1) at (9.5, 1.5) {0};
    \node[] (cf-1f) at (9.5, .5) {\small{$f-1$}};

\node[] (cfv0) at (12.5, 4.5) {1};
    \node[align=center,text width=1cm] (cfv1) at (12.5, 3.5) {\small{$n-f-1$}};
    \node[] (cfc0) at (12.5, 2.5) {0};
    \node[] (cfc1) at (12.5, 1.5) {0};
    \node[] (cff) at (12.5, .5) {\small{$f$}};

\node[] (c0f1v0) at (15.5, 4.5) {\small{$n-f$}};
    \node[] (cf1v1) at (15.5, 3.5) {0};
    \node[] (cf1c0) at (15.5, 2.5) {0};
    \node[] (cf1c1) at (15.5, 1.5) {0};
    \node[] (cf1f) at (15.5, .5) {\small{$f$}};

        \draw[->] (1, 3.5) to node[sloped,above] {\small{$r_3 \times (n-2)$}} (3, 3.5);
        \draw[->] (1, 3.5) to node[sloped,below] {\small{$r_5 \times 1$}} (3, 2.5);
        \draw[->] (1, 2.5) to node[sloped,below, yshift=.1] {\small{$r_7 \times 1$}} (3, .5);

        \draw[->] (4, 3.5) to node[sloped,above] {\small{$r_3 \times (n-3)$}} (6, 3.5);
        \draw[->] (4, 3.5) to node[sloped,below] {\small{$r_5 \times 1$}} (6, 2.5);
        \draw[->] (4, 2.5) to node[sloped,below, yshift=.2] {\small{$r_7 \times 1$}} (6, .5);
        \draw[->] (4, .5) to node[sloped,below] {\small{$r_9 \times 1$}} (6, .5);

        \node[] at (8.05, 2.65) {$\mathbf{\vdots}$};

    \draw[thick, ->] (7.05, 2.5) to (7.85, 2.5);

    \draw[thick, ->] (8.25, 2.5) to (8.95, 2.5);

        \draw[->] (10, 3.5) to node[sloped,above] {\small{$r_2 \times 1$}} (12, 4.5);
        \draw[->] (10, 3.5) to node[sloped,below,align=center,text width=1.9cm] {\small{$r_3 \times (n-f-1)$}} (12, 3.5);
        \draw[->] (10, 2.5) to node[sloped,below] {\small{$r_7 \times 1$}} (12, .5);
        \draw[->] (10, .5) to node[sloped,below] {\small{$r_9 \times (f-1)$}} (12, .5);

        \draw[->] (13, 4.5) to node[sloped,above] {\small{$r_1 \times 1$}} (15, 4.5);
        \draw[->] (13, 3.5) to node[sloped,below,align=center,text width=1.9cm] {\small{$r_2 \times (n-f-1)$}} (15, 4.5);
        \draw[->] (13, .5) to node[sloped,below] {\small{$r_9 \times f$}} (15, .5);

\end{tikzpicture}
}      \caption{An execution of \floodmin\ of length $f +1$.}%
     \label{fig:fm-long}
\end{figure}
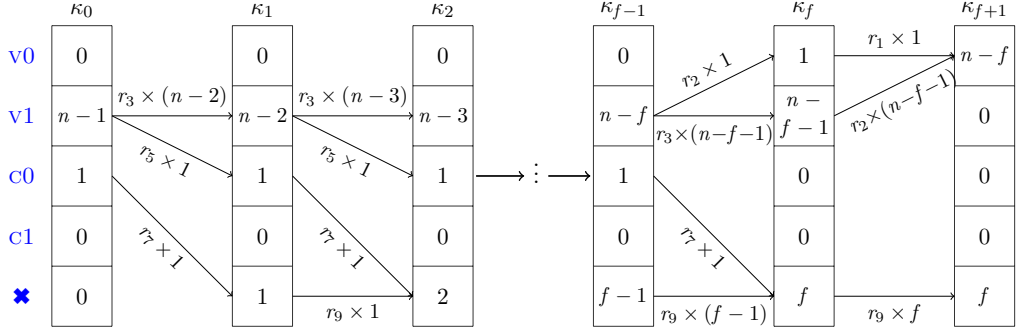

\begin{figure}
     \scalebox{.8}{
\begin{tikzpicture}

    \node[] (conf0) at (.5, 5.25) {$\kappa'_0$};
    \draw[step=1,black!80, very thin] (0, 0) grid (1, 5);

    \node[] (conf1) at (3.5, 5.25) {$\kappa'_1$};
    \draw[step=1,black!80, very thin] (3, 0) grid (4, 5);

    \node[] (conf2) at (6.5, 5.25) {$\kappa'_2$};
    \draw[step=1, black!80, very thin] (6, 0) grid (7, 5);

    \node[blue] (v0) at (-.5, 4.5) {$\VO$};
    \node[blue] (v1) at (-.5, 3.5) {$\VI$};
    \node[blue] (c0) at (-.5, 2.5) {$\CRO$};
    \node[blue] (c1) at (-.5, 1.5) {$\CRI$};
    \node[blue] (f) at (-.5, .5) {\DD};

\node[] (c0v0) at (.5, 4.5) {0};
    \node[] (c0v1) at (.5, 3.5) {\small{$n-1$}};
    \node[] (c0c0) at (.5, 2.5) {1};
    \node[] (c0c1) at (.5, 1.5) {0};
    \node[] (c0f) at (.5, .5) {0};

\node[] (c1v0) at (3.5, 4.5) {0};
    \node[] (c1v1) at (3.5, 3.5) {\small{$n-f$}};
    \node[] (c1c0) at (3.5, 2.5) {\small{$f-1$}};
    \node[] (c1c1) at (3.5, 1.5) {0};
    \node[] (c1f) at (3.5, .5) {1};

\node[] (c2v0) at (6.5, 4.5) {\small{$n-f$}};
    \node[] (c2v1) at (6.5, 3.5) {0};
    \node[] (c2c0) at (6.5, 2.5) {0};
    \node[] (c2c1) at (6.5, 1.5) {0};
    \node[] (c2f) at (6.5, .5) {\small{$f$}};

    \draw[->] (1, 3.5) to node[sloped,above] {\small{$r_3 \times (n-f)$}} (3, 3.5);
    \draw[->] (1, 3.5) to node[sloped,below] {\small{$r_5 \times (f-1)$}} (3, 2.5);
    \draw[->] (1, 2.5) to node[sloped,below, yshift=.1] {\small{$r_7 \times 1$}} (3, .5);

    \draw[->] (4, 3.5) to node[sloped,below,align=center,text width=1.9cm] {\small{$r_2 \times (n-f)$}} (6, 4.5);
    \draw[->] (4, 2.5) to node[sloped,above] {\small{$r_7 \times (f-1)$}} (6, .5);
    \draw[->] (4, .5) to node[sloped,below] {\small{$r_9 \times 1$}} (6, .5);

\end{tikzpicture}
}      \caption{An execution of \floodmin\ of length $2$ that reaches the
     same configuration as the long execution in Figure~\ref{fig:fm-long}.}%
     \label{fig:fm-short}
\end{figure}
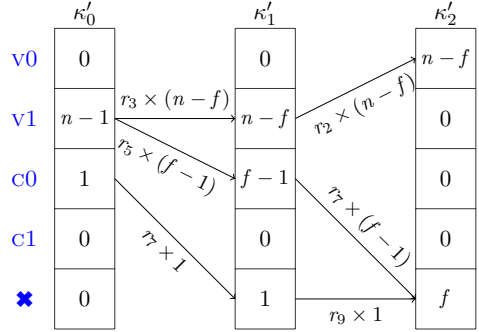

Consider again the example execution of  \floodmin\
depicted in Figure~\ref{fig:fm-long},
whose length (i.e., the number of transitions) is $f+1$.
That is, the length of this execution
is parameterized in the number $f$ of faults.
We see that after the first transition (which we described
above) we get a configuration where the values of the counters~$\kappa_1$
satisfy all guards in the STA again.
This is because there is more than one process in
the locations $\VO$ and $\CRO$ ($\kappa_1[\CRO] = 1$),
and no process in location $\VO$ ($\kappa_1[\VO] = 0$).
This means that in every round, we can repeat the same transition
where one process moves from
the location $\VI$ to the location $\CRO$, and the other processes
in location $\VI$ use the self-loop rule to stay in location $\VI$.
Hence, after $f-1$ transitions, the counters have the following values:
there are $n-f$ processes in location $\VI$, one process in
$\CRO$, and $f-1$ processes in \DD\ (see $\kappa_{f-1}$ in Figure~\ref{fig:fm-long}).
Since in $\kappa_{f-1}$ there are in total~$f$ processes in the locations reserved for the
faulty processes ($\kappa_{f-1}[\CRO] + \kappa_{f-1}[\mbox{\DD}] = f$),
no other processes can be moved
from $\VI$ to $\CRO$; instead, in the $f$-th transition,
one process moves from the location $\VI$ to $\VO$ using the rule $\arule_2$.
This makes the self-loop rule $\arule_3$ disabled, as its guard,
$\guard_2 \equiv \#\{\VO\} < 1$, evaluates to false.
This triggers the move of the $n-f-1$ process from
the location $\VI$ to the location $\VO$ in the
$f+1$-st (last) transition.

In Figure~\ref{fig:fm-short}, we see an execution of length
two, where in the first transition, $f-1$ processes move from the
location~$\VI$ to the location~$\CRO$ via the rule~$\arule_5$, and
in the second transition, $n-f$ processes move from the
location~$\VI$ to the location~$\VO$ using the rule~$\arule_2$.
That is, while a configuration where the counters
have values as in~$\kappa_{f+1}$ can be reached by a long
     execution as depicted in Figure~\ref{fig:fm-long},
     it can also be reached in just two steps
     as depicted in Figure~\ref{fig:fm-short}
     (observe that $\kappa'_2=\kappa_{f+1}$).
We are interested in whether there is a natural number~$k$
     (independent of the parameters $n$, $t$ and $f$)
     such that for an arbitrary execution in any counter system,
     we can always find a shorter execution starting and ending in the
     same configuration, such that the short execution is of
     length at most~$k$.
If this is the case, we say that the STA has \emph{bounded diameter}.
In~\cite{StoilkovskaKWZ19}, we formalize the concept of bounded
diameter for an STA, by adapting the
definition of diameter from~\cite{BiereCCZ99}.
We also introduce an SMT-based semi-decision procedure
for computing a bound on the diameter.
The procedure enumerates candidates for the diameter bound, and checks
     (by calling an SMT solver) if the number is indeed the
     diameter; if it finds such a bound, it  terminates.

\subsection{Bounded Model Checking}
To verify the safety properties of an algorithm for all
values of the parameters, we can solve the dual parameterized
reachability problem.
Namely,  a safety property holds for all values of the parameters
iff there do not exist values of the parameters for which
a bad configuration, where the safety property is violated,
is reachable.
The existence of a bound on the diameter motivates the use of
bounded model checking, as we can search for
violations of the a safety property in executions with length up to
     the diameter.
Crucially, this approach is complete: if an execution reaches a
     bad  configuration, this bad configuration is already reached by
     an execution of  bounded length.
Thus, once the diameter is found, we
     encode the violation of a safety property using a
     formula in Presburger
     arithmetic, and use an SMT solver to check for
     violations.
More details on how this encoding is done can be found
in~\cite{StoilkovskaKWZ19}.

The SMT queries that are used for computing the diameter and encoding the violation of the
safety properties contain quantifiers for dealing with the parameters symbolically.
Surprisingly, the performance of the SMT solvers on these queries is very good,
reflecting the recent progress in dealing with quantified queries.
We found that the diameter bounds of synchronous algorithms in the
literature are tiny (from 1 to 8), which makes our approach applicable in
practice.
The verified algorithms are given in Section~\ref{sec:bymc}.

\subsection{Undecidability}
In~\cite{StoilkovskaKWZ19}, we formalized the
parameterized reachability problem, which is the question
whether a certain configuration can be reached in some
counter system for some values of the parameters.
We showed that this problem is in general undecidable for STA,
by reduction from the halting problem of two-counter
machines~\cite{Minsky67}.
The detailed undecidability proof can be found in~\cite{SKWZ21:sttt}.
In particular, the undecidability implies that some STA
have unbounded diameters.
However, the SMT-based semi-decision procedure for computing
the diameter can still automatically compute the diameter bounds
for our benchmarks.
Remarkably, these bounds are independent of the parameters.
In addition, we also theoretically establish the existence
of a bound on the diameter, which is independent of the parameters,
for a class of STA that satisfy
certain conditions.

\section{Parameterized Verification of Asynchronous Algorithms}%
\label{sec:async}

\subsection{Asynchronous Algorithms}

In an asynchronous algorithm, at each time point,
exactly one process performs a step.
That is, the steps of the processes are interleaved.
This is in contrast to the synchronous semantics, where
all processes take steps simultaneously.
Given the pseudocode description of the algorithm, we can
interpret one line of code as an atomic
unit of execution of a process.
As in the synchronous case, the processes execute
send-to-all, receive, and local computation steps.
When a process performs a send-to-all step, it places
one message in the message buffers of all the other processes.
In the receive step, a process takes some messages out of the
     incoming message buffer.
Observe that not necessarily all messages that
were sent are in the buffer, and the process may possibly take
no message out of it.
The local computation step updates the local variables based on the
received messages.
Often, a step in the asynchronous semantics is more coarse-grained,
in the sense that it combines all three steps described above.
That is, in this coarse-grained case,
a step consists of receiving messages, updating the state, and
     sending one or more messages.

As we do not restrict which messages are taken out of the buffer
     during a step, we cannot bound the time needed for message
     transmission.
Moreover, we do not restrict the order in which processes  take steps,
     so we cannot bound the time between two steps of a single
     process.
Typically, we are interested in verifying safety (i.e., properties
     stating that nothing bad ever happens) under these conditions.
However, for liveness (i.e., properties stating that something  good
     eventually happens) this is problematic.
We need the messages to be delivered eventually,  and we need the
     correct processes to take steps from time to time.
That is, liveness is typically preconditioned by fairness guarantees:
     every correct processes takes infinitely many steps and every
     message sent from a correct process to a correct process is
     eventually received.
These constraints correspond roughly to the asynchronous system
     defined in~\cite{FLP85} in which consensus is not solvable.
However, these constraints are sufficient for fault-tolerant
     broadcast.
Intuitively, this is because broadcast has weaker liveness requirements
     than consensus.

\begin{figure}[t]
    \begin{minipage}{.55\linewidth}
\begin{lstlisting}[language=pseudo,columns=fullflexible,  numbers=left, numberstyle=\tiny,]
int v:=input(@$\{0, 1\}$@);
bool accept:=false;
while (true) do {
  if (v = 1) then send <ECHO> to all; @\label{st:sendecho}@
  receive messages from other processes;
  if received <ECHO> from $\ge$ t + 1 processes @\label{line:tp1}@
    then v:=1; @\label{line:v1}@
  if received <ECHO> from $\ge$ n - t processes@\label{line:tp2}@
    then accept:=true; @\label{line:acc}@
}
\end{lstlisting}
    \end{minipage}
    \hfill
    \begin{minipage}{.4\linewidth}

\tikzstyle{node}=[circle,draw=black,thick,minimum
size=3mm,font=\normalsize]
\tikzstyle{init}=[circle,draw=black!90,fill=green!10,thick,minimum
size=3mm,font=\normalsize]
\tikzstyle{final}=[circle,draw=black!90,fill=red!10,thick,minimum size=3mm,font=\normalsize]
\tikzstyle{rule}=[->,thick]
\tikzstyle{post}=[->,thick]
\tikzstyle{pre}=[<-,thick]
\tikzstyle{cond}=[rounded corners,rectangle,minimum
  width=1cm,draw=black,fill=white,font=\normalsize]
\tikzstyle{asign}=[rectangle,minimum
  width=1cm,draw=black,fill=gray!5,font=\normalsize]

\tikzset{every loop/.style={min distance=5mm,in=140,out=113,looseness=2}}
\resizebox{!}{.8\linewidth}{
\begin{tikzpicture}[>=latex]
 \node[] at (1.6,  2.25) [init,label=right:\textcolor{blue}{$\VO$}]
(1) {};
 \node[] at (-1.6, 2.25) [init,node,label=left:\textcolor{blue}{$\VI$}] (2) {};
 \node[] at (0, .5) [node,label=below left:\textcolor{blue}{$\SE$}] (3) {};
 \node[] at (0, -1) [final,label=below left:\textcolor{blue}{$\AC$}] (4) {};

 \draw[post] (2) edge[sloped]
         node[anchor=north, midway]
    {\small{$r_3$}}
    (3);

\draw[post, rounded corners=5pt] (1)
        |- node[anchor=north]
    {\small{$r_7$}}
    (4);

\draw[post] (1) --
        node[anchor=north, sloped, midway]
    {$r_2$} (3);

\draw[post, rounded corners=5pt] (2)
        |- node[anchor=north]
    {$r_8$}
    (4);

\draw[post] (3) to node[anchor=south, left]
    {$r_5$}
    (4);

\draw[rule] (1) to[out=150,in=210,looseness=8]
    node[anchor=east] {$r_1$} (1);
\draw[rule] (4) to[out=-30,in=-90,looseness=8]
    node[anchor=north] {$r_6$}  (4);
\draw[rule] (3) to[out=-25,in=25,looseness=8]
    node[anchor=west,near start] {$r_4$} (3);

\end{tikzpicture}
}
      \end{minipage}
    \caption{Pseudocode of reliable broadcast \`a la~\cite{ST87:abc} and its threshold automaton.}%
    \label{fig:ST}
\end{figure}
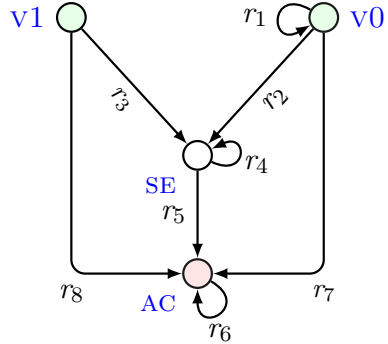

\begin{table}
    \caption{The rules of the TA from Figure~\ref{fig:ST}. We omit
    the rules $\arule_1, \arule_4, \arule_6$ as it has the trivial guard
    (\textit{true}) and no update.}%
\label{tab:strules}
    \small
\begin{center}
         \begin{tabular}{lll}
              \tbh{Rule} & \tbh{Guard} & \tbh{Update} \\
                   \hline
              $\arule_2$ & $\guard_1 \colon x \ge (t + 1) - f$ & $\cpp{x}$ \\
              $\arule_3$ & $\true$ & $\cpp{x}$\\
              $\arule_5$ & $\guard_2 \colon x \ge (n - t) - f $ & --- \\
              $\arule_7$ & $\guard_2 \colon x \ge (n - t) - f $ & $\cpp{x}$\\
              $\arule_8$ & $\guard_2 \colon x \ge (n - t) - f $ & $\cpp{x}$

         \end{tabular}
    \end{center}
    \end{table}

We now look at a
classic example of a fault-tolerant distributed algorithm, namely
the broadcasting algorithm by Srikanth and Toueg~\cite{ST87}, which
we refer to as~\strb.
The description of its code is given in Figure~\ref{fig:ST}.
The algorithm tolerates Byzantine failures,
where faulty processes may send messages only to a subset
of the processes (or even send conflicting
    data).
It solves the reliable broadcast problem, by ensuring that
every message sent to a non-faulty process is eventually
received.

The algorithm in Figure~\ref{fig:ST} does this by forwarding message
     content received from other processes and only accepting a
     message content if it was received from a quorum of processes.
One instance of the algorithm is executed for each  message
     content~$m$.
Initially, the variable~$\var{v}$ captures whether a process has
     received~$m$,  and, if this is the case, it stores the value~$1$.
A process that has received~$m$ (i.e., it has $\var{v}=1$)  sends
     \textsf{ECHO} to all (line~\ref{st:sendecho}).
In an implementation, the message would be of the form (\textsf{ECHO},
     $m$), that is, it would be tagged with \textsf{ECHO}, and carry
     the content~$m$ to distinguish different instances running in
     parallel; also it would suffice to send the message once instead
     of sending it in each iteration.
If the guard in line~\ref{line:tp1} evaluates to true at a
     process~$p$, then~$p$ has received~$t+1$ \textsf{ECHO} messages.
This means that at least one correct process has forwarded the
     message.
This triggers the process~$p$ to also forward the message.
If a process~$p$ receives $n-t$ \textsf{ECHO} messages, it finally
     accepts the  message content stored in~$m$ due to
     line~\ref{line:tp2}.
The reason this algorithm works is that the combination of  the
     thresholds $n-t$ and $t+1$, as well as the resilience condition
     $n>3t$, ensures that if one correct processes  has $n-t$
     \textsf{ECHO} messages, then every other correct process will
     eventually  receive at least $t+1$ messages  (there are $t+1$
     correct processes among any  $n-t$ processes).
This implies that every correct process will forward the message,
     and since there are at least $n-t$ correct processes, every
     correct process will accept.
However, manual reasoning about this arithmetics over parameters is
     subtle and error-prone.
Our verification techniques automate this reasoning; they are
     designed to deal efficiently with threshold expressions and
     resilience conditions.

\subsection{Asynchronous Threshold Automata}

Similarly as in STA, the nodes in \emph{asynchronous threshold
	automata (TA)} encode the locations of processes,
	and the edges (i.e., the rules) represent local transitions.
In addition, the TA
(i)~have shared variables, and
(ii)~their rules are labeled with expressions of the form
$\guard \mapsto \textsf{act}$.
A process moves along an edge labelled by $\guard \mapsto \textsf{act}$
	and performs an action~$\textsf{act}$,
	only if the condition $\guard$, called
	a \emph{threshold guard}, evaluates to true.

We model the algorithm \strb~\cite{ST87},
whose pseudocode is in Figure~\ref{fig:ST} on the left,
	using the TA shown in
	Figure~\ref{fig:ST} on the right.
We use a shared variable $x$ to capture the number of
	\textsf{ECHO} messages sent by correct processes.
We have two threshold guards:
$\guard_1 \colon x \ge (t+1) - f$ and
 	$\guard_2 \colon x \ge (n-t) - f$.
Depending on the initial value of a correct process, 0 or 1,
	the process is initially either in location $\VO$ or in $\VI$
        (which stand for ``value~0'' and ``value~1'', respectively).
If its value is~1, then the process broadcasts \textsf{ECHO},
	and executes the rule $r_3 \colon \true \mapsto \cpp{x}$.
This is modelled by a process moving from location~$\VI$ to
        location~$\SE$ (``sent echo'') and
	increasing the value of~$x$.
If its value is~0, then the process has to wait to receive enough
	messages.
	That is, it waits for $\guard_1$ to become true, and then it
	broadcasts the \textsf{ECHO}
	message and moves to location $\SE$.
This is captured by the rule $r_2$, which is labelled
by $\guard_1 \mapsto \cpp{x}$.
Finally, once a process has enough \textsf{ECHO}
	messages for the guard $\guard_2$ to evaluate to
    true, it sets accept to true and moves to~$\AC$ (``accept'': it
    encodes that the variable \textsf{accept} has been set to true;
    cf.\ line~\ref{line:acc} in Figure~\ref{fig:ST}).
Thus, the rule~$r_5$ is labelled by the guard~$\guard_2$ (and no update of the
shared variable~$x$),
whereas the rules~$r_7$ and~$r_8$ are labeled by $\guard_2\mapsto\cpp{x}$.

\subsection{Counter Systems}
Similarly to STA,
the semantics of TA is captured by counter systems.
Recall that in a counter system,
instead of storing the location of each process,
we count the number of
    processes in each location, as all processes are identical.
In the counter systems for the asynchronous
semantics, we also store the values of the
shared variables, which are incremented as
the processes execute the rules.
Therefore, a configuration in a counter system comprises:
(i)~the values of the counters for
    each location,
    (ii)~the values of the shared variables, and
    (iii)~the parameter values.
A configuration is initial if all processes are in initial locations (in our example,
	$\VO$ or $\VI$), and all shared variables have value~0 (in our example, $x=0$).
A transition of a process along a rule from location $\ell$ to
	location $\ell'$ --- labelled by $\guard \mapsto \textsf{act}$ ---
	is modelled by the following configuration update:
	(i)~the counter of $\ell$ is decreased by~1, and the counter of $\ell'$
	is increased by~1,
(ii)~the shared variables are updated according to the action $\textsf{act}$, and
	(iii)~the parameter values are unchanged.
Observe that this transition is possible only if the guard~$\guard$
of the rule evaluates to true in the original configuration.

As we will see below,
the key ingredient of our technique is acceleration of transitions, that is,
    we allow many processes to move along the same edge simultaneously.
In the resulting configuration,
the counters and shared variables are updated depending
	on the number of processes that participate
    in the accelerated transition.
It is important to notice that any accelerated transition
can be encoded in SMT\@.

\subsection{Reachability}%
\label{sec:asycnhreach}

In~\cite{KLVW17:FMSD}, we determine a finite set of execution ``patterns'',
    and then analyse each pattern separately.
These patterns restrict the order in which the threshold guards
	become true (if ever).
Namely, we observe how the set of guards that evaluate to true
	changes along each execution.
In our example TA for the algorithm \strb, given in Figure~\ref{fig:ST},
there are two (non-trivial) guards, $\guard_1$ and $\guard_2$.
Initially, both evaluate to false, as $x=0$.
During an exe\-cution, none, one, or both of them become true.
Note that once they become true, they can never evaluate to false again, as the
	number of sent messages, stored in the shared variable~$x$, cannot decrease.
Thus, there is a finite set of execution patterns.

For instance, a pattern
    $\{\}\dots\{\guard_1\}\dots\{\guard_1,\guard_2\}$ captures all finite
    executions $\tau$ that can be represented as $\tau = \tau_1 \cdot t_1 \cdot
    \tau_2 \cdot t_2 \cdot \tau_3$, where $\tau_1,\tau_2,\tau_3$ are
    sub-executions of $\tau$, and $t_1$ and $t_2$ are transitions.
No threshold guard is satisfied in a configuration visited by $\tau_1$, and only
    the guard~$\guard_1$ is satisfied in all configurations visited by $\tau_2$.
Both guards are satisfied in all configurations visited by~$\tau_3$.
Observe that in this pattern,
the transitions~$t_1$ and~$t_2$ change the evaluation of the guards.
Another pattern is given by $\{\}\dots\{\guard_2\}\dots\{\guard_1,\guard_2\}$.
Here, the guard~$\guard_2$ is satisfied before the guard~$\guard_1$.
In the pattern $\{\}\dots\{\guard_1\}$, the guard~$\guard_2$ is never satisfied.

To perform verification, we have to analyse all execution patterns.
For each pattern, we construct a so-called \emph{schema}, defined as a sequence of
    accelerated transitions, whose free variables are the number of processes that
    execute the transitions and the parameter values.
Recall that for the TA in Figure~\ref{fig:ST},
the process transitions are modelled by rules denoted with $r_i$, for
    $i \in \{1, \dots, 8\}$.
For instance, the pattern $\{\}\dots\{\guard_1\}$ produces the schema:

\begin{equation*}
    \mathcal{S} = \{\} \; \underbracket{r_1, r_3}_{\tau_1},
    \underbracket{r_3}_{t_1} \;\{ \guard_1 \}\; \underbracket{r_1, r_2, r_3,
    r_4}_{\tau_2} \;\{ \guard_1 \}\;.
\end{equation*}

The above schema has two segments, $\tau_1$ and
    $\tau_2$, corresponding to $\{\}$ and $\{\guard_1\}$, respectively.
In each of them we list all the rules that can be executed according to the
    guards that evaluate to true, in a fixed natural order: only $r_1$ and $r_3$ can be executed
    if no guard evaluates to true,
    and the rules $r_1, r_2, r_3, r_4$ can be executed only if the
    guard~$\guard_1$ holds true.
Additionally, we have to list all the candidate rules for the transition~$t_1$
    that can change
    the evaluation of the guards.
In our example, the guard~$\guard_1$ can evaluate to true only if
a transition along the rule~$r_3$ is taken.

We say that an execution follows the schema $\mathcal{S}$ if its transitions
    appear in the same order as in $\mathcal{S}$, but they are accelerated
    (every transition is executed by a number of processes, possibly zero).
For example, if $(r,k)$ denotes that~$k$ processes execute the rule~$r$
    simultaneously, then the execution $\rho = (r_1,2)(r_3,3)(r_2,2)(r_4,1)$
    follows the schema~$\mathcal{S}$, where the transitions of the form $(r,0)$ are
    omitted.
In this case, we prove that for each execution~$\tau$ of pattern
    $\{\}\dots\{\guard_1\}$, there
    is an execution~$\tau'$ that follows the schema~$\mathcal{S}$,
    and the executions~$\tau$ and~$\tau'$ reach the same configuration (when executed from
    the same initial configuration).
This is achieved by \emph{mover analysis}: inside any segment in which the set
    of guards that evaluate to true is fixed,
    we can swap adjacent transitions (that are not in
    a natural order).
In this way, we gather all transitions of the same rule
    next to each other, and transform them into a single accelerated
    transition.
For example, the execution $\tau = (r_3,2)(r_1,1)(r_3,1)(r_1,1)(r_2,1)(r_4,1)(r_2,1)$ can be
    transformed into the execution $\tau'= \rho$ from above, and they reach the same
    configurations.
Therefore, instead of checking reachability for all executions of the pattern
    $\{\}\dots\{\guard_1\}$, it is sufficient to analyse reachability
    only for the executions that
    follow the schema~$\mathcal{S}$.

Every schema is encoded as an SMT query over linear integer arithmetic with free
    variables for acceleration factors, parameters, and counters.
An SMT model gives us an execution of the counter system, which
    typically disproves safety.

For example, consider the following reachability problem: Can the system reach
    a configuration with at least one process in $\ell_3$?
For each SMT query, we add the constraint that the counter of $\ell_3$ is
    non-zero in the final configuration.
If the query is satisfiable, then there is an execution where at
    least one process reaches~$\ell_3$.
Otherwise, there is no such execution following the particular schema,
where a process reaches $\ell_3$.
That is why we have to check all schemas.

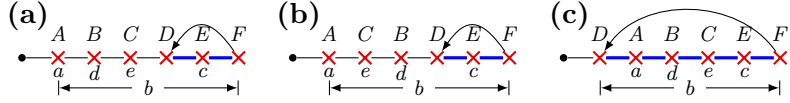
\begin{figure}[t]
    \centering
        \begin{tikzpicture}[x=1cm,y=1cm,font=\scriptsize,>=latex, scale=0.8];
    \tikzstyle{node}=[circle,fill=black,minimum size=0.1cm,inner sep=0cm];
    \tikzstyle{cut}=[cross out,thick,draw=red!90!black,
        minimum size=0.15cm,inner sep=0mm,outer sep=.1mm];
    \tikzstyle{path}=[-];
    \tikzstyle{Gfin}=[-, very thick, blue];

  \begin{scope}[xshift=0cm, yshift=0cm]
    \node at (0.1, .7) { \normalsize\textbf{(a)} };

    \node[node] (0) at (0, 0) {};

\foreach \x/\n in {.6/A, 1.2/B, 1.8/C, 2.4/D, 3.0/E, 3.6/F}
        \node[cut,label={above:$\n$}] (\n) at (\x,0) {};

    \draw[path] (0) -- (A);
    \draw[path] (A) -- (B);
    \draw[path] (B) -- (C);
    \draw[path] (C) -- (D);
    \draw[Gfin] (D) -- (E);
    \draw[Gfin] (E) -- (F);
    \draw[->] (F) edge[bend right=55,looseness=1.7] (D);

    \draw[|<->|] ($(A)+(0,-.5)$)
        --node[midway, fill=white, text=black]
        {$b$} ($(F)+(0,-.5)$);
    \node at ($(A)+(0,-.25)$) {$a$};
    \node at ($(B)+(0,-.25)$) {$d$};
    \node at ($(C)+(0,-.25)$) {$e$};
    \node at ($(E)+(0,-.25)$) {$c$};
  \end{scope}

  \begin{scope}[xshift=4.5cm, yshift=0cm]
    \node at (0.1, .7) { \normalsize\textbf{(b)} };

    \node[node] (0) at (0, 0) {};

\foreach \x/\n in {.6/A, 1.8/B, 1.2/C, 2.4/D, 3.0/E, 3.6/F}
        \node[cut,label={above:$\n$}] (\n) at (\x,0) {};

    \draw[path] (0) -- (A);
    \draw[path] (A) -- (C);
    \draw[path] (C) -- (B);
    \draw[path] (B) -- (D);
    \draw[Gfin] (D) -- (E);
    \draw[Gfin] (E) -- (F);
    \draw[->] (F) edge[bend right=55,looseness=1.7] (D);

    \draw[|<->|] ($(A)+(0,-.5)$)
        --node[midway, fill=white, text=black]
        {$b$} ($(F)+(0,-.5)$);
    \node at ($(A)+(0,-.25)$) {$a$};
    \node at ($(B)+(0,-.25)$) {$d$};
    \node at ($(C)+(0,-.25)$) {$e$};
    \node at ($(E)+(0,-.25)$) {$c$};
  \end{scope}

  \begin{scope}[xshift=9cm, yshift=0cm]
    \node at (0.1, .7) { \normalsize\textbf{(c)} };

    \node[node] (0) at (0, 0) {};

\foreach \x/\n in {1.2/A, 1.8/B, 2.4/C, .6/D, 3.0/E, 3.6/F}
        \node[cut,label={above:$\n$}] (\n) at (\x,0) {};

    \draw[path] (0) -- (D);
    \draw[Gfin] (D) -- (A);
    \draw[Gfin] (A) -- (B);
    \draw[Gfin] (B) -- (C);
    \draw[Gfin] (C) -- (E);
    \draw[Gfin] (E) -- (F);
    \draw[->] (F) edge[bend right=55] (D);

    \draw[|<->|] ($(D)+(0,-.5)$)
        --node[midway, fill=white, text=black]
        {$b$} ($(F)+(0,-.5)$);
    \node at ($(A)+(0,-.25)$) {$a$};
    \node at ($(B)+(0,-.25)$) {$d$};
    \node at ($(C)+(0,-.25)$) {$e$};
    \node at ($(E)+(0,-.25)$) {$c$};
  \end{scope}

\end{tikzpicture}

     \caption{Three out of 18 shapes of lassos that satisfy
            the formula $\ltlF (a \wedge \ltlF d \wedge \ltlF e
        \wedge \ltlG b \wedge \ltlG \ltlF c)$. The crosses show cut points for:
    (A)~formula $\ltlF (a \wedge \ltlF d \wedge \ltlF e
        \wedge \ltlG b \wedge \ltlG \ltlF c)$,
            (B)~formula $\ltlF d$, (C)~formula $\ltlF e$,
            (D)~loop start, (E)~formula~$\ltlF c$, and
            (F)~loop end.
    }%
    \label{fig:lasso-shapes}
\end{figure}

\subsection{Safety and Liveness}

In~\cite{KLVW17:POPL} we introduced a fragment of Linear Temporal Logic
    called~\ELTLTB{}.
Its atomic propositions test location counters for zero.
Moreover, this fragment uses only two temporal operators: $\ltlF$
    (eventually) and $\ltlG$ (globally).
Our goal is to check whether there exists a counterexample to a temporal
    property, and thus formulas in this fragment represent negations of safety
    and liveness properties.

Our technique for verification of safety and liveness properties uses the
	reachability method from Section~\ref{sec:asycnhreach} as its basis.
As before, we want to construct schemas that we can translate to SMT queries
	and check their satisfiability.
Note that violations of liveness properties are infinite executions of
    a lasso shape, that is, $\tau\cdot\rho^\omega$, where $\tau$ and
	$\rho$ are finite executions.
Hence, we have to enumerate the patterns of lassos.
These shapes depend not only on the values of the thresholds, but also on the
	evaluations of atomic propositions that appear in temporal properties.
We single out configurations in which atomic propositions evaluate to true, and
	call them \emph{cut points}, as they ``cut'' an execution into finitely
	many segments (see Figure~\ref{fig:lasso-shapes}).

We combine these cut points with those ``cuts'' in which the threshold
     guards become enabled (as in the reachability analysis).
All the possible orderings in which the evaluations of threshold
     guards and formulas become true, give us a finite set of lasso
     patterns.

We construct a schema for each shape by first defining schemas for
     each of the segments between two adjacent cut points.
On one hand, for reachability, it is sufficient to execute all enabled
     rules of that segment exactly once in the natural order.
Thus, each sub-execution $\tau_i$ can be transformed into~$\tau_i'$
     that follows the segment's schema, so that $\tau_i$ and $\tau_i'$
     reach the same final configuration.
On the other hand, the safety and liveness properties reason about
     atomic propositions inside executions.
To this end, we introduced a \emph{property-specific mover analysis}:
     While classic mover analysis~\cite{Lipton75} generates an alternative trace that
     leads to the same final configuration as the original trace,
     in property-specific mover
     analysis, we need to maintain not only the final configuration but
     also trace properties, e.g., invariants (that is,
     properties that hold in all configurations along the trace, and
     not only in the final one).
This new mover analysis allows us to construct schemas by executing
     all enabled rules a fixed number of times in a specific order.
The number of rule repetitions depends on a temporal property; it is
     typically two or three.

For each lasso pattern we encode its schema
	in SMT and check its satisfiability.
As~\ELTLTB{} formulas are negations of specifications,
	an~SMT model gives us a counterexample.
If no schema is satisfiable, the temporal property holds~true.

\section{Parameterized Verification of Asynchronous Randomized Multi-Round Algorithms}%
\label{sec:multi}

\subsection{Randomized Algorithms}\label{sec:RARARA}

\begin{figure}[t]
\begin{lstlisting}[language=pseudo,columns=fullflexible,multicols=2,  numbers=left, numberstyle=\tiny,]
bool v := input_value({0, 1});
int rnd := 1;
while (true) do
 send (R,rnd,v) to all;
 wait for n - t messages (R,rnd,*); @\label{ln:Rwait}@
 if received (n + t) / 2
   messages (R,rnd,w)
 then send (P,rnd,w,D) to all; @\label{ln:sendPval}@
 else send (P,rnd,?) to all;
 wait for n - t messages (P,rnd,*); @\label{ln:Pwait}@
 if received at least t + 1 @\label{ln:startbranch}@
    messages (P,rnd,w,D) then {
  v := w;
  if received at least (n + t) / 2@\label{ln:maj}@
   messages (P,rnd,w,D)
  then decide w;
 } else v := random({0, 1});@\label{ln:toincoss}@
 r := r + 1;
od @\label{ln:endbranch}@




\end{lstlisting}
    \caption{Pseudocode of the \benor\ algorithm for Byzantine faults, with $n>5t$}%
    \label{fig:bobpseudo}
\end{figure}

The randomized multi-round algorithms circumvent the impossibility
of asynchronous consensus~\cite{FLP85} by relaxing the termination requirement to
almost-sure termination, i.e., termination with
probability~1.
The processes execute an infinite sequence of asynchronous rounds, and
update their local variables based on the messages received only
in the current round.
Asynchronous algorithms with this feature are called
     \emph{communication closed}~\cite{EF82,DDMW19:CAV}.

A prominent example of a randomized multi-round algorithm  is
Ben-Or's fault-tolerant
     binary consensus~\cite{Ben-Or83} algorithm  in
     Figure~\ref{fig:bobpseudo}, which we refer to as~\benor.
While the algorithm is executed under the asynchronous semantics, the
     processes have a local variable~$\var{rnd}$ that stores the round number,
     and use it to tag the messages that they send in round~$\var{rnd}$.
Observe that the guards only take into account messages
which are tagged with the round number~$\var{rnd}$, that is, the
algorithm operates only on messages from the current
     round.
In the case of the \benor\ algorithm,
each round consists of two stages:
(i)~the processes first exchange messages
     tagged with $\var{R}$, wait until the number of received messages reaches a
     certain threshold (the expression over the parameters in
     line~\ref{ln:Rwait}),
(ii)~and then exchange messages tagged with~$\var{P}$.
As in the previous examples, $n$ denotes the number of processes, among
     which at most~$t$ may be faulty.
In the \benor\ algorithm in Figure~\ref{fig:bobpseudo}, the processes are
Byzantine-faulty.
The thresholds $n-t$, $(n+t)/2$ and $t+1$ that occur in the pseudocode,
combined with the
     resilience condition  $n > 5t$, ensure  that no two correct
     processes ever decide on different values.
If there is no ``strong majority'' for a decision value in line~\ref{ln:maj}, a
     process chooses a new value by tossing a coin in line~\ref{ln:toincoss}.

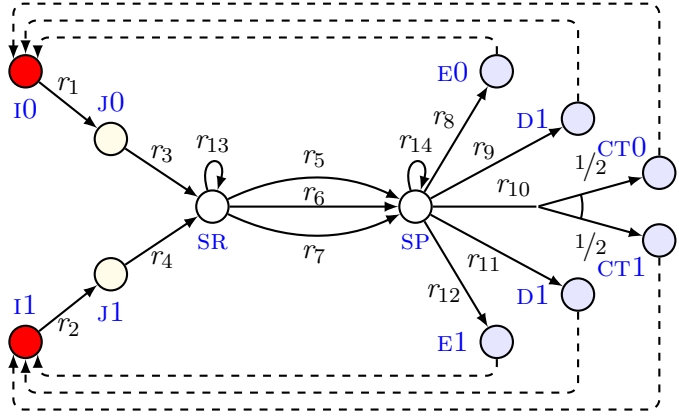
\begin{figure*}[t]
\begin{center}

\tikzstyle{node}=[circle,draw=black,thick,minimum size=4.3mm,inner sep=0.75mm,font=\normalsize]
\tikzstyle{init}=[node,fill=yellow!10]
\tikzstyle{final}=[node,fill=blue!10]
\tikzstyle{border}=[node, fill=red]
\tikzstyle{rule}=[->,thick]
\tikzstyle{post}=[->,thick,rounded corners,font=\normalsize]
\tikzstyle{postdash}=[->,thick,dashed,rounded corners,font=\normalsize]
\tikzstyle{pre}=[<-,thick]
\tikzstyle{cond}=[rounded
  corners,rectangle,minimum
  width=1cm,draw=black,fill=white,font=\normalsize]
\tikzstyle{asign}=[rectangle,minimum
  width=1cm,draw=black,fill=gray!5,font=\normalsize]

\tikzset{every loop/.style={min distance=5mm,in=140,out=113,looseness=2}}
\begin{tikzpicture}[>=latex, thick,scale=0.9, every node/.style={scale=01}, xscale=1.2]

 \node[] at (0.2, 2) [border,label=below:\textcolor{blue}{$\IO$}]    (0) {};
 \node[] at (0.2, -2) [border,label=above:\textcolor{blue}{$\II$}]   (1) {};

 \node[] at (1.25, 1) [init,label=above:\textcolor{blue}{$\JO$}]  (2) {};
 \node[] at (1.25, -1) [init,label=below:\textcolor{blue}{$\JI$}](3) {};

 \node[] at (2.5, 0) [node, label=below:\textcolor{blue}{$\SR$}] (4) {};

 \node[] at (5, 0) [node, label=below:\textcolor{blue}{$\SP$}] (9) {};

 \node[] at (6, 2) [final,label=left:\textcolor{blue}{$\EO$}]  (5) {};
 \node[] at (8, .5) [final,label={[label distance=-1.5mm]105:\textcolor{blue}{$\CTO$}}](6) {};
 \node[] at (8, -0.5) [final,label={[label distance=-1.5mm]255:\textcolor{blue}{$\CTI$}}](7) {};
 \node[] at (7, -1.3) [final,label=left:\textcolor{blue}{$\DI$}](8) {};
\node[] at (7, 1.3) [final,label=left:\textcolor{blue}{$\DI$}](11) {};
\node[] at (6, -2) [final,label=left:\textcolor{blue}{$\EI$}](12) {};

\node[inner sep=0,minimum size=0] at (6.5,0) (ct) {};

\draw[post] (0) to[] node[anchor=south] {$r_1$} (2); \draw[post] (1) to[] node[anchor=north] {$r_2$} (3);

\draw[post] (2) to[] node[anchor=south] {$r_3$} (4);
\draw[post] (3) to[] node[anchor=north] {$r_4$} (4);

\draw[post] (9) to[] node[anchor=south,xshift=-1ex]
    {$r_8$} (5);

\draw[post] (9) to[] node[anchor=south,pos=.4]
    {$r_9$} (11);

\draw[post] (9) to[] node[anchor=north,xshift=-1ex]
    {$r_{12}$} (12);

\draw[post] (9) to[] node[anchor=north,pos=.4]
    {$r_{11}$} (8);

\draw[post, bend right] (4) to[] node[anchor=north]
    {$r_7$} (9);

\draw[post, bend left] (4) to[] node[anchor=south]
    {$r_5 $} (9);

\draw[post] (4) to[] node[anchor=south,yshift=-.7ex]
    {$r_6$ } (9);

\draw[rule] (4) to[out=105,in=75,looseness=13] node[anchor=south]
    {$r_{13}$}
	     (4);

\draw[rule] (9) to[out=105,in=75,looseness=13] node[anchor=south]
    {$r_{14}$}
	     (9);

 \draw[-] (9) to[] node[anchor=south west]
    {$r_{10}$} (ct);
 \draw[post] (ct) to[]node[align=center,anchor=south, midway]
    {$\nicefrac{1}{2}$} (6);
 \draw[post] (ct) to[]node[align=center,anchor=north, midway]
    {$\nicefrac{1}{2}$} (7);
 \path pic[draw, angle radius=6mm,angle eccentricity=1.2] {angle = 7--ct--6};

\draw[postdash] (5.north) -- ($(5)+(0,.5)$) -| (0.north east);

\draw[postdash] (6.north) -- ($(6)+(0,2.5)$) -| (0.north west);
 \draw[postdash] (11.north) -- ($(11)+(0,1.45)$) -| (0.north);

 \draw[postdash] (7.south) -- ($(7)+(0,-2.5)$) -| (1.south west);
 \draw[postdash] (12.south) -- ($(12)+(0,-.5)$) -| (1.south east);\draw[postdash] (8.south) -- ($(8)+(0,-1.45)$) -| (1.south);

\end{tikzpicture}
 \end{center}
\caption{The \benor\ alorithm as PTA with resilience condition $n>3t
  \wedge t>0 \wedge t \ge f\ge 0$.}%
  \label{fig:ex-pta}
\end{figure*}

\begin{table*}
\caption{The rules of the PTA from Figure~\ref{fig:ex-pta}. We omit
rules $r_1, r_2, r_{13}, r_{14}$ as they have the trivial guard
(\textit{true}) and no update. }
\small
\begin{center}
    \begin{tabular}{lll}
        \tbh{Rule} & \tbh{Guard} & \tbh{Update} \\
            \hline
$r_3$ & $\mathit{true}$ & $\cpp{x_0}$ \\
            $r_4$ & $\mathit{true}$ & $\cpp{x_1}$ \\
        $r_5$ & $x_0{+}x_1 \ge n{-}t{-}f \quad \wedge \quad x_0 \ge (n{+}t)/2 -f$
            & $\cpp{y_0}$ \\
        $r_6$ & $x_0{+}x_1 \ge n{-}t{-}f \quad \wedge \quad x_1 \ge (n{+}t)/2 -f$
            & $\cpp{y_1}$ \\
        $r_7$ & $x_0{+}x_1 \ge n{-}t{-}f \quad \wedge \quad x_0 \ge
        (n{-}3t)/2 -f \quad \wedge \quad \ x_1 \ge (n{-}3t)/2 -f$ & $\cpp{y_?}$ \\
        $r_8$ &
     $y_0{+}y_1{+}y_? \ge n{-}t{-}f \quad \wedge \quad y_? \ge (n{-}3t)/2 -f
     \quad \wedge\quad y_0 \ge  t{+}1{-}f$ & ---\\
     $r_9$ & $y_0{+}y_1{+}y_? \ge n{-}t{-}f \quad\wedge\quad y_0 > (n{+}t)/2 -f$ & ---\\
     $r_{10}$ & $y_0{+}y_1{+}y_? \ge n{-}t{-}f \quad\wedge\quad y_? \ge (n{-}3t)/2-f
     \quad\wedge\quad y_? > n{-}2t{-}f{-}1$ & --- \\
     $r_{11}$ & $y_0{+}y_1{+}y_? \ge n{-}t{-}f \quad\wedge\quad y_1 > (n{+}t)/2 -f$ & ---\\
     $r_{12}$ & $y_0{+}y_1{+}y_? \ge n{-}t{-}f \quad \wedge \quad y_? \ge (n{-}3t)/2 -f
     \quad \wedge\quad y_1 \ge  t{+}1{-}f$ & ---\end{tabular}%
    \label{tab:pta-rules}
  \end{center}
\end{table*}

\subsection{Probabilistic Threshold Automata}

As discussed above,
randomized algorithms typically have an unbounded number of asynchronous rounds
	and randomized choices.
Probabilistic threshold automata (PTA), introduced in~\cite{bertrand_et_al19},
are extensions of asynchronous TA
	that allow formalizing these features.

A PTA modelling the \benor\ algorithm from Figure~\ref{fig:bobpseudo}
	is shown in Figure~\ref{fig:ex-pta}.
The behavior of a process in a single round is modelled by the solid edges.
Note that in this case, the threshold guards should be evaluated according
	to the values of shared variables, e.g., $x_0$ and $x_1$, in the observed round.
The dashed edges model round switches: once a process reaches a final location in
	a round, it moves to an initial location of the next round.
The coin toss is modelled by the branching rule $r_{10}$: a process in location $\SP$
	can reach either location~$\CTO$ or location~$\CTI$ by moving along this fork,
	both with probability $1/2$.

\subsection{Unboundedly Many Rounds}\label{sec:properprope}

In order to overcome the issue of unboundedly many rounds, we prove that we can
    verify PTA by ana\-lysing a one-round automaton that fits in the framework
    of Section~\ref{sec:async}.
In~\cite{bertrand_et_al19}, we prove that one can reorder the transitions of any
    fair execution such that their round numbers are in {a non-decreasing}
    order.
The obtained ordered execution is stutter-equivalent to the original one.
Thus, both executions satisfy the same~\LTLX\ properties over the
    atomic propositions of one round.
In other words, the distributed system can be transformed to a
    sequential composition of one-round systems.

The main problem with isolating a one-round system is that the consensus
	specifications often talk about at least two different rounds.
In this case we need to use round invariants that imply the specifications.
For example, if we want to verify agreement, we have to check that
	no two processes decide different values, possibly in different
	rounds. We do this in two steps: (i)~we check the round invariant
	that no process changes its decision from round to round, and
	(ii)~we check that within a round no two processes disagree.

\subsection{Probabilistic Properties}\label{sec:probaprope}

The semantics of a PTA is an
     infinite-state Markov decision process (MDP), where the
     non-determinism is traditionally resolved  by an adversary.
In~\cite{bertrand_et_al19}, we restrict our attention to so-called
     \emph{round-rigid adversaries}, that is, fair adversaries  that
     generate executions in which a process enters round~$\var{rnd}+1$
     only after all processes finished round~$\var{rnd}$.
We introduced these adversaries as a first step towards parameterized
     verification of fault-tolerant randomized distributed algorithms
     for probabilistic properties.
Intuitively, in the probabilistic world, round-rigid adversaries are
     the correspondence of representative executions in the
     non-deterministic models that are the
     result of a reduction for communication-closed
     algorithms~\cite{EF82,Chaouch-SaadCM09,DDMW19:CAV}.
In Section~\ref{sec:weak} we discuss the link between round-rigid
     adversaries and the standard notion of weak adversaries.

Verifying almost-sure termination under round-rigid adversaries	calls for distinct arguments.
Our methodology follows the lines of the manual proof of Ben-Or's
	consensus algorithm by Aguilera and Toueg~\cite{AT12}.
However, our arguments are not specific to Ben-Or's algorithm,
	and apply to other randomized distributed algorithms
	(see Table~\ref{fig:bymc-history}).
Compared to their paper-and-pencil proof, the threshold automata framework
  required us to provide a more formal setting and a more
  informative proof, also pinpointing the needed hypotheses.
The crucial parts of our proof are automatically checked by the model
  checker ByMC\@.

\subsection{Weak Adversaries}\label{sec:weak}

The approach from  Section~\ref{sec:probaprope} leaves a gap between
     round-rigid adversaries and the classic adversary definitions we
     find in the distributed computing literature.
This problem is addressed in~\cite{VMCAI42} where the standard notion
     of a ``weak adversary'' is considered.
Weak adversaries pose a formalization challenge in the counter system
     semantics of TA\@.
The reason is that these adversaries are defined over individual
     processes and messages; notions that do not exist in the counter
     system representation.
As a result, a more concrete semantics of TA was
     introduced, which explicitly captures processes, sets of received
     messages for each process, and threshold guards over the number
     of specific messages in these sets.
For this semantics,~\cite{VMCAI42} contains a reduction theorem from
     weak adversaries to round-rigid adversaries.
While~\cite{VMCAI42} does not contain a formalization of the
     abstraction step from the explicit model to TA,
     we conjecture that such a proof can be done based on
     the ideas that underlie Section~\ref{sec:pseudoSTA} below.
     If this is the case, the verification results from
     Section~\ref{sec:multi}  and the lower part of
     Table~\ref{fig:bymc-history} hold under a wider  class of
     adversaries.

\section{Modeling}%
\label{sec:modeling}

\subsection{From Pseudocode to Threshold Automata}\label{sec:pseudoSTA}
Observe that the parameterized verification approaches,  presented in
     Sections~\ref{sec:syncTA},~\ref{sec:async}, and~\ref{sec:multi},
     take as input threshold automata, whose guards are evaluated
     over the global state (i.e., the messages sent by the processes).
When modeling threshold-guarded distributed algorithms with
     verification in mind, we are faced with a  formalization gap
     between the threshold automata and the algorithm descriptions
     given in terms of pseudocode, which is supposed to run locally
     on a node and contains guards over the local state (i.e.,
     the messages received locally by a process).
For many cases, this formalization gap is easy to overcome,  i.e., the
     translation from pseudocode to a threshold automaton  is
     immediate, and can easily be done manually.
However, for some algorithms this is not the case.

Consider the \benor\ algorithm given in
     Figure~\ref{fig:bobpseudo}.
The main challenge we faced in the formalization is to express the
     path that leads to the coin toss in line~\ref{ln:toincoss}.
Observe that to reach line~\ref{ln:toincoss},
a process has received at least $n-t$ messages with the label $\var{P}$
(line~\ref{ln:Pwait}), as well as
less than $t+1$ messages of type $\var{(P, rnd, 0, D)}$ and
less than $t+1$ messages of type $\var{(P, rnd, 1, D)}$,
since the placeholder $\var{w}$ in line~\ref{ln:startbranch} can take both input values from the
set $\{0, 1\}$.
In other words, if we were to rewrite the pseudocode as a set of
     guarded commands, the coin toss would be guarded by an
     expression containing
     (i)~a lower bound condition over the local variables, i.e., the
     number of received messages labeled by $\var{P}$ (at least $n-t$ messages), and
     (ii)~two upper bound conditions over the local variables, i.e., the number of
     received messages of type $\var{(P, rnd, w, D)}$ (less than $t+1$ messages),
     where $\var{w} \in \{0, 1\}$.
However, as is required by the threshold automata framework,
the constraints over the global variables that can lead to the coin
     toss are captured by the guard of the rule $r_{10}$, given in
     Figure~\ref{fig:ex-pta}.
It only contains lower bound
     conditions over the global variables, which are the number of
     globally sent messages.
This translation between lower and upper bound conditions expressed
over the local and global variables is non-trivial, and when done manually, requires
     intuition on the operation of the algorithm.
As a result, in writing our benchmarks, we observed that when done
     manually, this translation is error-prone.

\newcommand{\rcvta}{\textsf{rcvTA}} \newcommand{\sndta}{\textsf{sndTA}} \newcommand{\ns}[1]{\mathsf{ns}({#1})}
\newcommand{\nr}[2]{\mathsf{nr}_{#1}(#2)}
\newcommand{\nsf}[1]{\mathsf{ns}_f({#1})}
\newcommand{\msgtype}{m}
\newcommand{\msgtypeset}{M}
\newcommand{\proci}{i}
\newcommand{\invariant}{\mathsf{Env}} \newcommand{\abs}[1]{\widehat{#1}}

In~\cite{Stoilkovska0WZ20,VMCAI41}, we address the problem of automating the
translation from pseudocode to a threshold automaton, for both
asynchronous and synchronous threshold-guarded distributed algorithms.
For randomized algorithms, whose local control flow
motivated this line of work, the same
results as for the asynchronous algorithms apply.
In order to automate the translation, we need to formalize the local
     transition relation expressed by the pseudocode.
To this end, we introduce a variant of TA, called
\emph{receive TA},
whose rules are guarded by expressions  over the
\emph{local} receive variables.

Let $\nr{\proci}{\msgtype}$ denote such a \emph{receive variable}, which
     encodes how many messages of type $\msgtype$ process~$\proci$ has
     received, and let  $\ns{\msgtype}$ denote a \emph{send variable},
     which stores the number of sent messages of that type.
Translating guards over receive variables $\nr{\proci}{\msgtype}$  to guards over
     send variables $\ns{\msgtype}$, for each message type~$\msgtype$, is based on
     \emph{quantifier elimination} for Presburger
     arithmetic~\cite{presburger1929vollstandigkeit,cooper1972theorem,pugh1992practical}.
In order to obtain the most precise guards over the send variables,
in the quantifier elimination step, the guards over the receive
variables are strengthened by an
\emph{environment assumption}~$\invariant$.
As we will see in Section~\ref{sec:faults} below, the environment assumption
 encodes the relationship between the receive and send variables, which
depends on the degree of synchrony and the fault model.
Given a guard $\guard$ over the receive variables,  a guard
     $\abs{\guard}$ over the send variables can be computed
     automatically by applying quantifier elimination to the formula
     $\guard' \equiv \exists \nr{\proci}{\msgtype_0} \dots
     \exists \nr{\proci}{\msgtype_k}\; (\guard \wedge \invariant)$,
     where $\msgtype_0, \dots, \msgtype_k$ are the message types
     that define the messages exchanged in the execution of the algorithm.
This produces a quantifier-free formula $\abs{\guard}$ over the send
     variables.

The translation procedure was implemented in a prototype~\cite{Stoilkovska0WZ20,VMCAI41}
that automatically generates guards over
     the send variables, by using Z3~\cite{MouraB08} to automate the
     quantifier elimination step.
By applying the translation procedure to the guard of every rule in the
receive threshold automaton given as input,
a threshold automaton with no receive variables is obtained automatically.
In~\cite{Stoilkovska0WZ20,VMCAI41}, it was shown that the translation
procedure based on quantifier elimination
is sound for both the asynchronous and synchronous case.
This means that a system of $n$ copies of an automatically generated
TA over the send variables is an \emph{overapproximation}
of a system of  $n$ copies of the receive TA given as input.
For a class of distributed algorithms that captures typical
     distributed algorithms found in the literature, it was also shown to be complete.

The translation procedure based on quantifier elimination thus
closes the formalization gap between the original description  of
     an algorithm (using received messages) and the  threshold
     automaton of the algorithm, given as an input to a
     verification tool.
More precisely, parameterized verification of threshold-guarded
distributed algorithms, starting with a formal model of the pseudocode
given by a receive threshold automaton,
can be fully automated by:
(i)~automatically producing a formal model suitable for verification by
applying the translation procedure based on quantifier elimination
and (ii)~automatically verifying its correctness by applying existing tools.

\subsection{Modeling Faults in Threshold Automata}%
\label{sec:faults}

To model the behavior that the faults introduce, when producing a
     (receive) TA for  a given algorithm, we have to
     capture the semantics of executing the code on a faulty process.
To capture the faulty semantics in the  automaton, we typically need to
     introduce additional locations or rules, depending on the fault
     model.
Also, depending on the fault model, we have different constraints  on
     the values of the receive and send variables, which are encoded
     by  an \emph{environment assumption}~$\invariant$ (required to
     obtain the more precise guards after the quantifier elimination
     discussed above).

We consider two types of faults in this paper -- crash and Byzantine
     faults.
In the case of crash faults, a process may crash in the middle of a
     send-to-all operation, which results in the message being sent
     only to a subset of processes.
In the case of Byzantine faults, no assumptions are made on the
     internal behavior of the faulty processes.
That is, Byzantine-faulty processes may send any message in any order
     to any process or fail to send messages.

\subsubsection{Crash Faults}%
\label{sec:crashfaults}
Crash-faulty processes stop executing the algorithm prematurely and cannot restart.
To model this behavior in a TA, we add so-called ``crash'' locations to which
processes move from the ``correct'' locations.
Processes that move to the ``crash'' locations remain there forever.
In addition, we introduce send variables $\nsf{\msgtype}$,
for each message type~$\msgtype$,
that count the number of messages of type~$\msgtype$ sent
by processes which are \emph{crashing}, i.e., by processes that are moving from
a ``correct'' to a ``crashed'' location.

The TA thus models the behavior of both correct  and
     faulty processes explicitly.
This allows us to express so-called \emph{uniform} properties that also refer
to states of faulty processes.\footnote{For instance, in consensus
  ``uniform agreement states that ``no two processes decide on
  different values'', while (non-uniform) ``agreement'' states that
  ``no two \emph{correct} processes decide on
  different values''.}
The environment assumption~$\invariant$ imposes
     constraints on the  number of processes allowed to populate the
     ``crash'' locations, and on  the number of received messages of
     each message type~$\msgtype$.
In particular, the environment assumption~$\invariant$ requires that there
     are  at most~$f$ processes in the ``crash'' locations, where~$f$
     is the  number of faulty processes.
Additionally, for each message type~$\msgtype$, the number of received
     messages  of type~$\msgtype$, stored in the receive
     variable~$\nr{\proci}{\msgtype}$ does  not exceed the number of
     messages of type~$\msgtype$ sent by the correct and  the
     crash-faulty processes, stored in the send variables
     $\ns{\msgtype}$ and  $\nsf{\msgtype}$, respectively.

In the synchronous case, where there exist strict guarantees on the
message delivery, the environment assumption~$\invariant$ also
bounds the value of the receive variables from below:
the constraint $\ns{\msgtype} \le \nr{\proci}{\msgtype}$
encodes that all messages sent by correct processes are received in the
round in which they are sent.

\subsubsection{Byzantine Faults}%
\label{sec:byzfaults}

To model the behavior of the Byzantine-faulty processes, which can act
     arbitrarily, no new locations and rules are introduced in the
     TA\@.
Instead, the TA is used to model the behavior of the
     correct processes, and the effect that the Byzantine-faulty
     processes have on the correct ones is captured in the guards and
     the environment assumption.
The number of messages sent by the Byzantine-faulty processes is
     overapproximated by the parameter~$f$, which denotes the number
     of faults.
That is, in the environment assumption~$\invariant$, we have the
     constraint $\nr{\proci}{\msgtype} \le \ns{\msgtype} + f$, which
     captures that the number of received messages of type~$\msgtype$
     does not exceed $\ns{\msgtype} + f$, which is an upper bound on
     the number of messages sent by the correct and Byzantine-faulty
     processes.

Since we do not introduce locations that explicitly model
the behavior of the Byzantine-faulty processes,
the TA is used to model the behavior of the $n-f$ correct
processes only.\footnote{As classically no assumptions are made on the
       internals of Byzantine processes, it does not make sense to
       consider uniform properties. Thus we also do not need Byzantine
     faults explicit in the model.}
In addition to the constraint that bounds the number of received messages
from above, the environment assumption for
the synchronous case also contains the constraint~$\ns{\msgtype} \le \nr{\proci}{\msgtype}$.
It is used to bound the number of received messages from below, and
ensure that all messages sent by correct processes are received.

\section{Tools: Byzantine Model Checker and SyncTA}\label{sec:bymc}

Byzantine Model Checker (ByMC) is an experimental tool for verification of
    threshold-guarded distributed algorithms.
Version~2.4.4 is available from GitHub~\cite{bymc2015:online}.
It implements the techniques for systems of asynchronous threshold automata and
    probabilistic threshold automata, which are discussed in
Section~\ref{sec:async} and Section~\ref{sec:multi}.
ByMC is implemented in OCaml and it integrates with SMT solvers via the SMTLIB
    interface~\cite{BarFT-SMTLIB} as well as OCaml bindings for Microsoft
    Z3~\cite{MouraB08}.

The techniques for synchronous threshold automata---discussed in
    Section~\ref{sec:techniques}---are implemented in another prototype
    tool~\cite{STAexperiments}.
This prototype is implemented in Python and it integrates with the SMT solvers
    Microsoft Z3 and CVC4~\cite{BarrettCDHJKRT11} via the SMTLIB interface.
The reason for this prototype not being integrated in ByMC is technical:
    Synchronous threshold automata required another parser and another internal
    representation.

\subsection{Inputs to ByMC}

ByMC accepts two kinds of inputs: (asynchronous) threshold automata written in
    the tool-specific format and threshold-guarded algorithms written in
    Parametric Promela~\cite{JKSVW13:SPIN}.
We do not give complete examples here, as the both formats are quite verbose.
A discussion about the both formats can be found in the tool paper~\cite{KW18}.

To give the reader an idea about the inputs we presents excerpts of the
    reliable broadcast, which was compactly presented using the graphical
    notation in Figure~\ref{fig:ST}.

All benchmarks discussed in this paper are available from the open access
    repository~\cite{benchmarks}.

\begin{figure}
\begin{lstlisting}[language=promela,numbers=left,frame=single,
    numberstyle=\scriptsize, columns=fullflexible]
#define V0      0 /* the initial state */
#define V1      1 /* the init message received */
#define SE      2 /* the echo message sent */
#define AC      3 /* the accepting state */
#define PC_SZ   4

symbolic int N; /* the number of processes: correct + faulty */
symbolic int T; /* the threshold */
symbolic int F; /* the actual number of faulty processes */

int nsnt = 0;   /* the number of messages sent by the correct processes */

assume(N > 3 * T && T > 0 && F >= 0 && F <= T);

atomic prec_unforg = all(Proc:pc == V0);
atomic prec_corr = all(Proc:pc == V1);
atomic ex_acc = some(Proc:pc == AC);
atomic all_acc = all(Proc:pc == AC);
atomic in_transit = some(Proc:nrcvd < nsnt);

active[N - F] proctype Proc() {
    byte pc = 0; byte next_pc = 0; /* the program counter (and its next value) */
    int nrcvd = 0; int next_nrcvd = 0; /* the number of received messages (and the next value) */
    /* non-deterministically choose the initial value between 0 and 1 */
    if ::pc = V0;
      ::pc = V1;
    fi;
end:
    do
       :: atomic {
            /* receive messages: pick a value between nrcvd and nsnt + F */
            havoc(next_nrcvd); assume(nrcvd <= next_nrcvd && next_nrcvd <= nsnt + F);
          /* a step by FSM */
          if /* find the next value of the program counter */
            :: next_nrcvd >= N - T -> next_pc = AC;
            :: next_nrcvd < N - T && (pc == V1 || next_nrcvd >= T + 1) -> next_pc = SE;
            :: else -> next_pc = pc;
          fi;
          if /* send the echo message */
            :: (pc == V0 || pc == V1) && (next_pc == SE || next_pc == AC) -> nsnt++;
            :: else;
          fi;
          /* update the variables to hold the next values */
          pc = next_pc; nrcvd = next_nrcvd;
       }
    od;
}

ltl fairness { []<>(!in_transit) }
ltl relay { ([](ex_acc -> <>(all_acc))) }
ltl corr { (prec_corr -> <>(ex_acc)) }
ltl unforg { (prec_unforg -> [](!ex_acc)) }
\end{lstlisting}
    \caption{Excerpt of reliable broadcast in Parametric Promela}%
    \label{fig:strb-promela}
\end{figure}

\subsection{Overview of the techniques implemented in ByMC}

\begin{table}[t]
    \caption{Asynchronous and randomized fault-tolerant distributed
    algorithms that are verified by different generations of ByMC\@.
For every technique and algorithm we show, whether the technique could verify
    the properties: safety (S), liveness (L), almost-sure termination under
    round-rigid adversaries (RRT), or none of them (-).}
   \begin{center}
   \begin{tabular}{l|ccchhh}
        & \tbh{SMT-S}
        & \tbh{SMT-L}
        & \tbh{SMT+MR}
        & \tbh{CA+SPIN}
        & \tbh{CA+BDD}
        & \tbh{CA+SAT}
        \\
        & {\small\cite{KVW15:CAV}}
        & {\small\cite{KLVW17:POPL}}
        & {\small\cite{bertrand_et_al19}}
        & {\small\cite{JohnKSVW13:fmcad}}
        & {\small\cite{KVW17:IandC}}
        & {\small\cite{KVW17:IandC}}
        \\
        \tbh{Algorithm}
        & {\small\cite{JohnKSVW13:fmcad}}
        & {\small\cite{KVW17:IandC}}
        & {\small\cite{KVW17:IandC}}
        & {\small\cite{KVW15:CAV}}
        & {\small\cite{KLVW17:POPL}}
        & {\small\cite{bertrand_et_al19}}
        \\\hline

    \tbh{FRB}~{\small\cite{CT96}}
        & S
        & S+L
        & --
        & S+L
        & S+L
        & S
        \\
    \tbh{STRB}~{\small\cite{ST87:abc}}
        & S
        & S+L
        & --
        & S+L
        & S+L
        & S
        \\
    \tbh{ABA}~{\small\cite{BrachaT85}}
        & S
        & S+L
        & --
        & --
        & S+L
        & --
        \\
    \tbh{NBACG}~{\small\cite{Gue02}}
        & S
        & S+L
        & --
        & --
        & --
        & --
        \\
    \tbh{NBACR}~{\small\cite{Raynal97}}
        & S
        & S+L
        & --
        & --
        & --
        & --
        \\
    \tbh{CBC}~{\small\cite{MostefaouiMPR03}}
        & S
        & S+L
        & --
        & --
        & --
        & --
        \\
    \tbh{CF1S}~{\small\cite{DobreS06}}
        & S
        & S+L
        & --
        & --
        & S+L
        & --
        \\
    \tbh{C1CS}~{\small\cite{BrasileiroGMR01}}
        & S
        & S+L
        & --
        & --
        & --
        & --
        \\
    \tbh{BOSCO}~{\small\cite{SongR08}}
        & S
        & S+L
        & --
        & --
        & --
        & --
        \\
        \hline
    \tbh{Ben-Or}~{\small\cite{Ben-Or83}}
        & --
        & --
        & S+RRT
        & --
        & --
        & --
        \\
    \tbh{RABC}~{\small\cite{Bracha87}}
        & --
        & --
        & S+RRT
        & --
        & --
        & --
        \\
    \tbh{kSet}~{\small\cite{MostefaouiMR18}}
        & --
        & --
        & S+RRT
        & --
        & --
        & --
        \\
    \tbh{RS-BOSCO}
        & --
        & --
        & S+RRT
        & --
        & --
        & --
        \\
    {\small\cite{SongR08}}
        &
        &
        &
        &
        &
        &
        \\
        \hline
   \end{tabular}
   \end{center}%
    \label{fig:bymc-history}
\end{table}

Table~\ref{fig:bymc-history} shows coverage of various asynchronous
    and randomized algorithms with these techniques.

The most performant SMT-based techniques in Table~\ref{fig:bymc-history} are
presented in this paper and match its sections as follows:

\begin{itemize}
    \item The technique for safety verification of asynchronous algorithms
        is called \textsf{SMT-S} (see Section~\ref{sec:async}).

    \item The technique for liveness verification of asynchronous algorithms
        is called \textsf{SMT-L} (see Section~\ref{sec:async}).

    \item The technique for verification of multi-round randomized algorithms is called
        \textsf{SMT-MR} (see Section~\ref{sec:multi}).

\end{itemize}

\noindent
For comparison, we give figures for the older techniques~\textsf{CA+SPIN},
    \textsf{CA+BDD}, and \textsf{CA+SAT}, which are shown in
    \colorbox{gray!30}{gray}.
For a detailed tutorial on these older techniques, we refer the reader
    to~\cite{KVW16:psi}.
For the context, we give only a brief introduction to these techniques below.
As one can see, the techniques that are highlighted in this survey are the most
efficient techniques that are implemented in ByMC\@.
After the brief introduction of the techniques, we give further experimental
    explanation for the efficiency of the SMT-based techniques.

\paragraph{Technique \textsf{CA+SPIN}}

This was the first technique that was implemented in ByMC in
    2012~\cite{JohnKSVW13:fmcad,2012arXiv1210.3847J,GKSVW14:SFM}.
In contrast to the SMT-based approaches highlighted in this survey, this
    technique abstracts every integer counter $\counters[\ell]$ to an abstract
    value that corresponds to a symbolic interval.
The set of symbolic intervals is constructed automatically by static analysis
    of threshold guards and temporal properties.
For instance, the reliable broadcast in Figure~\ref{fig:ST} would have four
    abstract values that correspond to the intervals: $\ropen{0, 1}$, $\ropen{1, t+1}$,
    $\ropen{t+1, n-t}$, and $\ropen{n-t, \infty}$.
This finite-state counter abstraction is automatically constructed from a
    protocol specification in Parameterized Promela and is automatically
    checked with the classical finite-state model checker Spin~\cite{H2003}.
As this abstraction was typically too coarse for liveness checking, we have
    implemented a simple counterexample-guided abstraction refinement loop for
    parameterized systems.
This approach extends the classic $\{0,1,\infty\}$-counter abstraction,
    which was designed for parameterized verification of mutual exclusion
    algorithms~\cite{PXZ02}.
Although this technique is a nice combination of classical ideas, we could
    only verify two broadcast algorithms by running it.
Not surprisingly, Spin was either running out of memory or by timeout.

\paragraph{Technique \textsf{CA+BDD}}

This technique implements the same counter abstraction as in \textsf{CA+SPIN},
    but runs the symbolic model checker nuXmv instead of Spin, which implements
    model checking with BDDs and SAT solvers~\cite{CavadaCDGMMMRT14}.
Although this extension scaled better than \textsf{CA+SPIN}, we could only
    check two additional benchmarks by running it.
Detailed discussions of the techniques \textsf{CA+SPIN} and \textsf{CA+BDD} can
    be found in~\cite{GKSVW14:SFM,KVW16:psi}.

\paragraph{Technique \textsf{CA+SAT}}

By running the abstraction/checking loop in nuXmv, we found that the bounded
    model checking algorithms of nuXmv could check long executions of our
     benchmarks.
However, bounded model checking in general does not have completeness
    guarantees.
In~\cite{KVW14:concur,KVW17:IandC}, we have shown that the counter systems of
     (asynchronous) threshold automata have computable bounded diameters, which
    gave us a way to use bounded model checking as a complete verification
    approach for reachability properties.
Still, the computed upper bounds were too high for achieving complete
    verification.

\subsection{Comparing the benchmark sizes.} 
\begin{figure}
    \includegraphics[width=.85\textwidth]{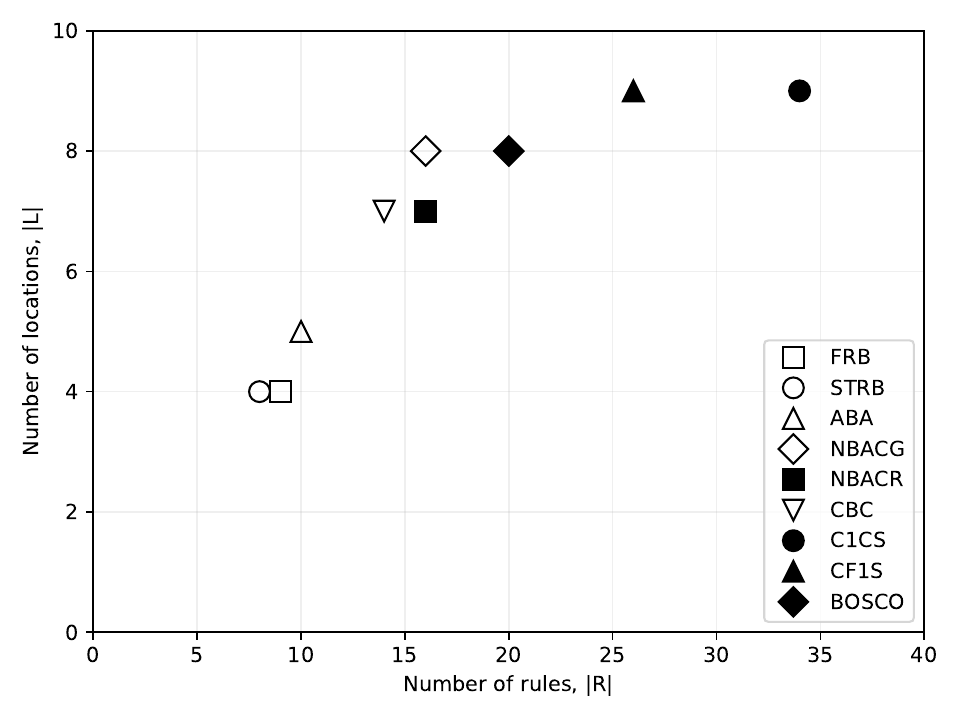}
    \caption{Relative sizes of the threshold automata used as input}%
    \label{fig:ta-sizes}
\end{figure}

\begin{figure}
    \includegraphics[width=.85\textwidth]{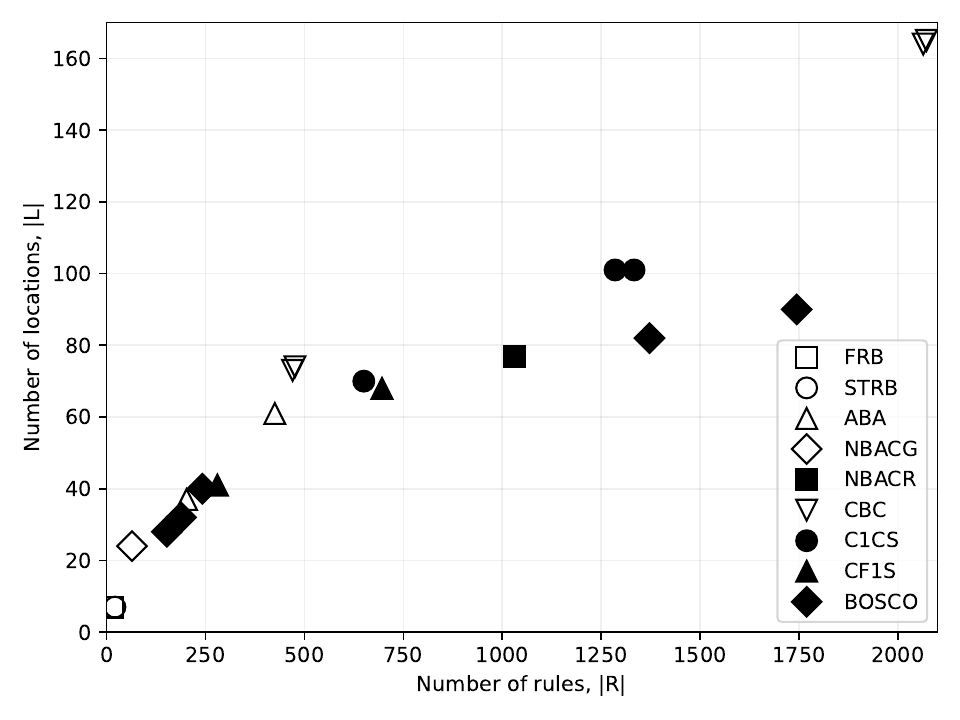}
    \caption{Relative size of the benchmarks in parametric Promela, after applying counter abstraction}%
    \label{fig:ca-sizes}
\end{figure}

To explain the performance of the
    techniques that belong to the \texttt{SMT} and \texttt{CA} groups, we
    present the sizes of the benchmarks in Table~\ref{tab:benchmark-sizes}.
Plots~\ref{fig:ta-sizes} and~\ref{fig:ca-sizes} visualize the number of
    locations and rules in threshold automata and the result of the counter
    abstraction of Parametric Promela, respectively.
As one can see, the threshold automata produced by counter abstraction are
    significantly larger in the number of locations and rules than the
    threshold automata over integer counters.
The techniques based on counter abstraction require finite domains.
Hence, the explosion in the number of locations and rules is a natural result
of applying abstraction.
In contrast, by reasoning over integer counters in the SMT-based techniques we
    keep the size of the input relatively small.
In our understanding, this should explain better efficiency of the SMT-based
    approaches.

\begin{table}
    \centering
    \caption{The relative sizes of the benchmarks introduced in~\cite{KW18}.
    }\label{tab:benchmark-sizes}
\begin{tabular}{rlc|rrcc}\hline{}
          \tbh{\#}
        & \tbh{Input}
        & \tbh{Case}
        & \multicolumn{4}{c|}{\tbh{Threshold}}
          \\
        & \multicolumn{2}{c|}{}
        & \multicolumn{4}{c}{\tbh{Automaton}}
          \\
        & \tbh{FTDA} & \scriptsize{(if more than one)}
        & \tbh{$|\mathbf{\local}|$}
        & \tbh{$|\mathbf{\ruleset}|$}
        & \tbh{$|\PrecondU|$}
        & \tbh{$|\PrecondL|$}
        \begin{filecontents*}{isola-summary.csv}
input,case,nlocs,nrules,nunlocking,nlocking,nschemas,schema-len,seq-sec,mpi-avg-sec,mpi-max-sec,seq-mem,mpi-avg-mem,rtype
frb,-,7,14,1,0,5,34,1,-,-,0.1,-,
frb,\emph{hand-coded \TA},4,9,1,1,70,38,1,-,-,0.1,-,ta
strb,-,7,21,3,0,18,72,1,-,-,0.1,-,
strb,\emph{hand-coded \TA},4,8,2,0,38,22,1,-,-,0.1,-,ta
nbacg,-,24,64,4,0,90,243,6,-,-,0.1,-,
nbacg,\emph{hand-coded \TA},8,16,0,1,5,54,1,-,-,0.1,-,ta
nbacr,-,77,1031,6,0,517,2489,523,-,-,0.7,-,
nbacr,\emph{hand-coded \TA},7,16,0,1,18,63,1,-,-,0.1,-,ta
aba,$\frac{n+t}{2} = 2t+1$,37,202,6,0,1172,850,659,12,13,1.0,0.2,
aba,$\frac{n+t}{2} > 2t+1$,61,425,8,0,5204,2112,53992,1440,1442,7.2,0.6,
aba,\emph{hand-coded \TA},5,10,2,2,542,57,14,-,-,0.1,-,ta
cbc,$\lfloor\frac{n}{2}\rfloor < n-t \wedge f=0$,164,2064,0,0,2,8168,1603,290,290,9.3,0.2,
cbc,$\lfloor\frac{n}{2}\rfloor = n-t \wedge f=0$,73,470,0,0,2,1790,27,9,9,0.6,0.1,
cbc,$\lfloor\frac{n}{2}\rfloor < n-t \wedge f>0$,165,2072,0,1,4,10213,10024,4943,4943,18.8,0.5,
cbc,$\lfloor\frac{n}{2}\rfloor = n-t \wedge f>0$,74,476,0,1,4,2258,273,47,47,1.5,0.1,
cbc,\emph{hand-coded \TA},7,14,0,1,5,56,1,-,-,0.1,-,ta
cf1s,$f=0$,41,280,4,0,90,770,45,5,8,0.2,0.1,
cf1s,$f=1$,41,280,4,1,523,787,257,6,6,0.4,0.1,
cf1s,$f>1$,68,696,6,1,3429,2132,10346,29,29,3.8,0.2,
cf1s,\emph{hand-coded \TA},9,26,3,3,13700,122,687,6,8,2.1,0.1,ta
c1cs,$f=0$,101,1285,8,0,251,460,331,38,38,0.8,0.1,
c1cs,$f=1$,70,650,6,1,448,303,239,11,11,0.4,0.1,
c1cs,$f>1$,101,1333,8,1,2100,404,1865,89,89,1.3,0.4,
c1cs,\emph{hand-coded \TA},9,30,7,3,$3.2 \cdot 10^6$,$\approx 400$,\normalsize{\clock},979,981,17.3,1.6,ta
bosco,$\lfloor\frac{n+3t}{2}\rfloor+1 = n-t$,28,152,6,0,20,423,4,3,4,0.1,0.1,
bosco,$\lfloor\frac{n+3t}{2}\rfloor+1 > n-t$,40,242,8,0,70,1038,29,6,6,0.2,0.1,
bosco,$\lfloor\frac{n+3t}{2}\rfloor+1 < n-t$,32,188,6,0,20,476,4,4,4,0.1,0.1,
bosco,$n>5t \wedge f=0$,82,1372,12,0,3431,27,265,35,35,0.3,0.4,
bosco,$n>7t$,90,1744,12,0,3431,179,1325,52,52,1.0,0.6,
bosco,\emph{hand-coded \TA},8,20,3,4,3429,43,82,4,4,0.2,0.1,ta
\end{filecontents*}
\csvreader[late after line=\ifcsvprostrequal{\rtype}{ta}{\\\rowcolor{gray!20}}{\\},before first line=\\\hline,late after last line=\\\hline]{isola-summary.csv}{1=\algo, 2=\rc, 3=\nl, 4=\nr, 5=\gu, 6=\gl, 14=\rtype
    }{\scriptsize{\thecsvrow}
        & \scriptsize{\textsf{\textbf{{\algo}}}}
            & \scriptsize{\rc}
            & \scriptsize{\nl}
            & \scriptsize{\nr} & \scriptsize{\gu} & \scriptsize{\gl}
    }
   \end{tabular}

\end{table}

\subsection{Model checking synchronous threshold automata.}

The bounded model checking approach for STA introduced in
Section~\ref{sec:syncTA} is not yet integrated into ByMC\@.
It is implemented as a stand-alone tool, available at~\cite{STAexperiments}.
In~\cite{StoilkovskaKWZ19}, we encoded multiple synchronous
algorithms from the literature, such as
    consensus~\cite{Lyn96,2010Raynal,BermanGP89,KINGTR,BielySW11},
$k$-set agreement (from~\cite{Lyn96}, whose pseudocode is
given in Figure~\ref{fig:FloodMin} and~\cite{2010Raynal}),
and reliable broadcast (from~\cite{ST87:abc,BielySW11})
algorithms.
We use Z3~\cite{MouraB08} and CVC4~\cite{BarrettCDHJKRT11}
as back-end SMT solvers.
Table~\ref{tab:syncTA} gives an overview of the verified synchronous
algorithms.
For further details on the experimental results, see~\cite{StoilkovskaKWZ19}.

\begin{table}[t]
\caption{Synchronous fault-tolerant distributed algorithms
verified with
the bounded model checking approach from~\cite{StoilkovskaKWZ19}.
With $\checkmark$ we show that: the SMT based
procedure finds a diameter bound with Z3 (\tbh{DIAM+Z3})
and CVC4 (\tbh{DIAM+CVC4});
there is a theoretical bound on the diameter (\tbh{DIAM+THM}).
We verify safety (S) by
bounded model checking with Z3 (\tbh{BMC+Z3}) and CVC4 (\tbh{BMC+CVC4}).}
\begin{center}
\begin{tabular}{l|ccccc}
    \tbh{Algorithm} &~\tbh{DIAM+Z3}
               &~\tbh{DIAM+CVC4}
               &~\tbh{DIAM+THM}
               &~\tbh{BMC+Z3}
               &~\tbh{BMC+CVC4}
               \\
    \hline
    \tbh{\floodset}~\cite{Lyn96} & $\checkmark$ & $\checkmark$ & -- & S & S  \\
    \tbh{\faircons}~\cite{2010Raynal}  & $\checkmark$ & $\checkmark$ & -- & S & S \\
    \tbh{\phaseking}~\cite{BermanGP89} & $\checkmark$ & $\checkmark$ & -- & S & S  \\
    \tbh{\phasequeen}~\cite{KINGTR}  & $\checkmark$ &$\checkmark$ & -- & S & S \\
    \tbh{\hybridking}~\cite{BielySW11} &$\checkmark$ &$\checkmark$ & -- & S & S \\
    \tbh{\byzking}~\cite{BielySW11}  & $\checkmark$ &$\checkmark$ & -- & S & S\\
    \tbh{\omitking}~\cite{BielySW11}   & $\checkmark$ &$\checkmark$ & -- & S & S \\
    \tbh{\hybridqueen}~\cite{BielySW11} & -- & -- & -- & -- & -- \\
    \tbh{\byzqueen}~\cite{BielySW11} & $\checkmark$ &$\checkmark$ & -- & S & S\\
    \tbh{\omitqueen}~\cite{BielySW11} & $\checkmark$ &$\checkmark$ & -- & S & S\\
\hline

\tbh{\floodmin}~\cite{Lyn96} & $\checkmark$ &$\checkmark$ & --& S & S \\
\tbh{\floodminomit}~\cite{BielySW11} & $\checkmark$ &$\checkmark$ & -- & S & S\\
\tbh{\ksetomit}~\cite{2010Raynal}  & -- & -- & -- & -- & -- \\

\hline
\tbh{\strb}~\cite{ST87:abc} & $\checkmark$ &$\checkmark$ & $\checkmark$& S & S \\
\tbh{\hybridrb}~\cite{BielySW11} & $\checkmark$ &$\checkmark$ & $\checkmark$& S & S \\
\tbh{\omitrb}~\cite{BielySW11} & $\checkmark$ &$\checkmark$ & $\checkmark$& S & S \\
\hline
\end{tabular}
\end{center}%
\label{tab:syncTA}
\end{table}

\section{Towards verification of Tendermint consensus}\label{sec:Tendermint}

Tendermint consensus is a fault-tolerant distributed algorithm for
     proof-of-stake blockchains introduced in~\cite{buchman2018latest}.
Tendermint can handle Byzantine faults under the assumption of partial
     synchrony.
It is running in the Cosmos network, where currently over 100 validator nodes
     are committing transactions and are managing the ATOM
     cryptocurrency~\cite{buchman2018cosmos}.

\subsection{Challenges of verifying Tendermint}    Tendermint consensus
    heavily relies on threshold guards, as can be seen from its
    pseudocode in~\cite{buchman2018latest}[Algorithm 1].
For instance, one of the Tendermint rules has the following precondition:
\begin{small}
\begin{align}
\textbf{upon} &\left<\mathsf{PROPOSAL}, h_p, round_p, v, *\right>\textbf{ from }proposer(h_p, round_p) \notag{}\\
    &\textbf{ AND }
2f+1 \left<\mathsf{PREVOTE}, h_p, round_p, id(v)\right> \notag{}\\
    &\textbf{ while }valid(v) \land step_p \ge \mathsf{prevote} \text{ for the first time}%
    \label{eq:tendermint}
\end{align}
\end{small}
The rule~(\ref{eq:tendermint}) requires two kinds of messages: (1)~a
     single message of type \textsf{PROPOSAL} carrying a proposal~$v$
     from the process $proposer(h_p, round_p)$ that is identified by
     the current round~$round_p$ and consensus instance~$h_p$, and
     (2)~messages of type \textsf{PREVOTE} from several nodes.
Here the term $2F+1$ (taken from the original paper) in fact does not
     refer to a number of processes.
Rather, each process has a voting power (an integer that expresses how
     many votes a process has), and $2F+1$ (in combination with
     $N=3T+1$) expresses that nodes that have sent \textsf{PREVOTE}
     have more than two-thirds of the voting power.
Although this rule bears similarity with the rules of threshold
     automata, Tendermint consensus has the following features that
     cannot be directly modelled with threshold automata:

\begin{enumerate}
\item In every consensus instance~$h_p$ and round~$round_p$, a single proposer
    sends a value that the nodes vote on.
The identity of the proposer can be accessed with the function $proposer(h_p,
    round_p)$.
\emph{This feature breaks symmetry among individual nodes}, which is required by
    our modelling with counter systems.
Moreover, the proposer function should be fairly distributed among the nodes,
    e.g., it can be implemented with round robin.

\item Whereas the classical example algorithms in this paper count messages, Tendermint evaluates the voting power of the
    nodes from which messages were received.
    This adds an additional layer of complexity.

\item Liveness of Tendermint requires the distributed system to reach a
    global stabilization period, when every message could be delivered not later
    than after a bounded delay.
This model of partial synchrony lies between synchronous and asynchronous
    computations and requires novel techniques for parameterized verification.

\end{enumerate}

\subsection{Checking parameterized one-round safety with ByMC}

While we are not able to verify the complete Tendermint consensus
     algorithm in ByMC, we use ByMC to verify its one-round safety in
     the parameterized case.
We do this, first we manually construct a threshold automaton from the
     pseudo code.
We translate the above rule into the following rules of a threshold automaton
    (an explanation of the syntax can found in~\cite{KW18}):

\begin{verbatim}
      3: locPrevote -> locPrecommit
          when (nprop0 >= 1 && nprevote0 >= 2 * T + 1 - F)
          do {
            nprecommit0' == nprecommit0 + 1;
            nprecommitAll' == nprecommitAll + 1;
            unchanged(nprop0, nprop1,
                      nprevote0, nprevote1, nprevoteNil, nprevoteAll,
                      nprecommit1, nprecommitNil);
          };
      4: locPrevote -> locPrecommit
          when (nprop1 >= 1 && nprevote1 >= 2 * T + 1 - F)
          do {
            nprecommit1' == nprecommit1 + 1;
            nprecommitAll' == nprecommitAll + 1;
            unchanged(nprop0, nprop1,
                      nprevote0, nprevote1, nprevoteNil, nprevoteAll,
                      nprecommit0, nprecommitNil);
          };
\end{verbatim}

We encoded the complete threshold automaton. The encoding is publicly available at:
\url{https://github.com/konnov/fault-tolerant-benchmarks/tree/master/lmcs20}.

Before
discussing ByMC's performance on this benchmark, we quickly discuss an
alternative encoding of this benchmark in \tlap{} for comparison with a standard model checker.

\subsection{Tendermint in \texorpdfstring{\tlap{}}{TLA+}}

\tlap{} is a general specification
     language~\cite{lamport2002specifying}, so it is much easier to
     write the first specification in~\tlap{}, rather than to write
     down a threshold automaton right away.
Additionally, we first debugged our \tlap{} specification with the TLC
     model checker~\cite{yu1999model}.
The~\tlap{} specification can be found in Appendix~\ref{sec:TATLA}.
The following snippet encodes the rule~(\ref{eq:tendermint}) discussed
     above, and indeed is a threshold-guarded rule that encodes a
     transition from a process in location ``prevote'' to location
     ``precommit'', provided enough ``propose'' and ``prevote''
     messages are received:

\medskip

\begin{center}
  \includegraphics{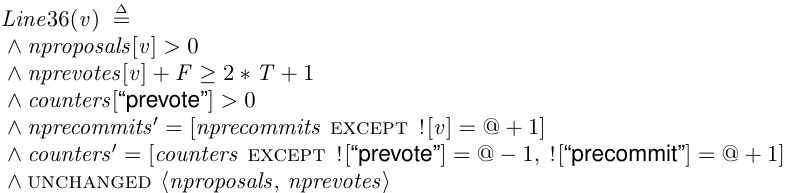}
\end{center}

\subsection{Comparison of verification performance}

We produced two artifacts, a threshold automaton and a TLA+ model\footnote{The \tlap{} specification
and the threshold automaton are publicly available at:
\url{https://github.com/konnov/fault-tolerant-benchmarks/tree/master/lmcs20}.}.

The standard~\tlap{} model checker TLC can check the specification
    for $N \le 20$ in reasonable time.
We checked the safety of consensus in one round: No two correct processes
     decide differently in the same round.
As TLC is an explicit-state model checker, it did
not scale to the value of
    $N = 125$, which is the current number of
     validators in the Cosmos main chain.
In contrast, we have verified this property with ByMC in the parameterized
    case in 2 seconds.

In this model of the algorithm, the one-round safety is satisfied under the
assumption of $N = 3 \cdot T + 1 \land T \ge F$. When we change the assumption
to either $N \ge 3 \cdot T + 1 \land T \ge F$, or $N = 3 \cdot T + 1$, ByMC
produces counterexamples to the property. The actual implementation of
Tendermint is different, as it dynamically recomputes the voting powers of
the validators and thus it recomputes the thresholds. However, we follow the
assumptions of~\cite{buchman2018latest} in our modeling.

We observe that our manual abstraction of the \tlap{} specification
into a threshold automaton is rather mechanical (while presented in
reverse order in this paper, we actually did the \tlap{} encoding first). This abstraction could be
done in a fully automatic pipeline: From \tlap{} to a receive threshold
automaton, and from a receive threshold automaton to a threshold automaton
    (see Section~\ref{sec:pseudoSTA}).
We are planning to use automatic abstractions to
     build a bridge between ByMC and Apalache --- a symbolic model checker
     for~\tlap{}~\cite{KT19}.

\subsection{Open problems for parameterized verification of multi-round safety}

To verify multi-round safety of Tendermint, we would like to invoke a reduction
argument similar to the one explained in Section~\ref{sec:properprope}.
However, Tendermint contains the following rule that prevents us from directly
applying the reduction result:

\begin{small}
\begin{align}
\textbf{upon} &\left<\mathsf{PROPOSAL}, h_p, round_p, v, vr\right>\textbf{ from }proposer(h_p, round_p) \notag{}\\
    &\textbf{ AND }
2f+1 \left<\mathsf{PREVOTE}, h_p, vr, id(v)\right> \notag{}\\
&\textbf{ while } step_p = \mathsf{propose} \land (vr \ge 0 \land vr < round_p)
    \notag{}\\
    &\dots%
    \label{eq:tendermint-past}
\end{align}
\end{small}

The rule in Equation~(\ref{eq:tendermint-past}) allows a process to make a step
by using messages from a past round~$vr$. As a result, Tendermint is not
communication-closed~\cite{EF82,Chaouch-SaadCM09,DDMW19:CAV}. Extending the
reduction argument to multi-round systems that are not communication closed
is subject to our ongoing work.

\section{Discussions and Related Work}%
\label{sec:related}

In our work, we follow the idea of identifying fragments of automata and logic
    that are sufficiently expressive for capturing interesting
    distributed algorithms and
    specifications, as well as amenable for completely automated
    verification.
We introduced threshold automata for that and implemented our verification
    techniques in the open source tool ByMC~\cite{KW18}.
By doing so, we verified several challenging distributed algorithms; most of
    them were verified for the first time.

Our work is in the context of computer-aided verification of
     distributed algorithms and systems, which  is a lively research
     area.
Parameterized verification of fault-tolerant distributed algorithms
has recently been addressed with a wide range of techniques,
most of which focus on asynchronous distributed algorithms.
In Section~\ref{sec:verifdistalg},
we survey both existing parameterized model checking techniques
and deductive verification techniques, based on producing
mechanized proofs using proof assistants.

Further, systems that implement fault-tolerant distributed
algorithms are very complex and increasingly hard to get right.
With increasing numbers of systems that implement
fault-tolerant distributed algorithms~\cite{Burrows06,JunqueiraRS11,Moraru:2013},
there is an interest in developing tool support for eliminating
flaws in distributed algorithms and their implementations by means of automated
verification.
We survey some of the approaches in this area in Section~\ref{sec:verifdistsys}.

Finally, the threshold automata framework has proved to be both of practical
     relevance as well as of theoretical interest.
There are several ongoing projects that consider complexity theoretic analysis of
     verification problems, as well as more refined probabilistic reasoning.
In Section~\ref{sec:relatedTA}, we give brief overview of these approaches.

\subsection{Verification of Distributed Algorithms}%
\label{sec:verifdistalg}
In this section, we survey existing model checking techniques, as well as
    deductive verification techniques, based on producing
    mechanized proofs using proof assistants.

Several consensus algorithms were verified for small system sizes
in~\cite{TsuchiyaS11,NoguchiTK12,DelzannoTT14}, by applying
model checking to the fixed size instances of up to, e.g., six processes.
In the parameterized case,
a framework for verifying fault-tolerance of distributed protocols
based on regular model checking was proposed in~\cite{FismanKL08}.
In this framework, the fault model specification is separate
from the process specification, which is similar our idea of
separating the process and environment specifications, and keeping
the system specification modular.
The approach was manually applied to verify the correctness of an
authenticated broadcast protocol that tolerates crash faults in the parameterized case.
Parameterized model checking of safety properties for fault-tolerant distributed algorithms using
counter abstractions and SMT solvers was proposed in~\cite{AlbertiGOP16}.
Using this approach, the authors automatically verified
two authenticated broadcast protocols, operating under different fault
models, namely crash, send-omission, general omission, and Byzantine faults.

The Heard-Of model~\cite{Charron-BostS09} was proposed as a
formalization framework for round-based, message passing distributed algorithms,
where the computation and fault models are captured by
so-called communication predicates.
This framework enables a systematic encoding of
(asynchronous, synchronous, or partially synchronous)
round-based algorithms and facilitates the comparison between different algorithms.
Several parameterized verification techniques
are designed for algorithms formalized in the heard-of model.
For partially synchronous consensus algorithms, expressed in an extension of the
heard-of model,~\cite{DHVWZ14} introduced a
consensus logic and (semi-)decision procedures for verifying user-provided
invariants.
In~\cite{MaricSB17}, a characterization of partially synchronous consensus
algorithms in the heard-of model was given. Based on this characterization,
the authors proved cut-off theorems specialized to the properties of consensus:
agreement, validity, and termination.
More recently,~\cite{GanjeiREP20}
proposed an approach for parameterized verification of safety properties of
round-based algorithms, that can be expressed in the heard-of model,
by combining overapproximation and backward reachability analysis.

Many fault-tolerant distributed algorithms have been
formalized using the specification language TLA+~\cite{lamport2002specifying},
e.g.,~\cite{GafniL03,Lamport11a,Moraru:2013}.
As TLA+ is equipped with an explicit state model checker, TLC~\cite{YuML99},
and a proof system, TLAPS~\cite{ChaudhuriDLM10}, often TLC is used
to debug the TLA+ specification of a given algorithm, for small and
fixed system sizes, and show the correctness of the algorithm in the parameterized
case by writing a machine-checkable proof in TLAPS\@.
In~\cite{TranKW20}, a cutoff result for failure-detection algorithms
was presented, the algorithms from~\cite{ChandraT96} was specified using TLA+,
and
TLC and APALACHE~\cite{KT19}, a new symbolic model checker for
TLA+, were used to verify its correctness for the cutoff size.
Other theorem provers have also been used to certify the
correctness of fault-tolerant distributed algorithms.
PVS~\cite{OwreRS92} was used in~\cite{LR93}, where a bug in an already
published synchronous consensus algorithm
tolerating hybrid faults was reported, and in~\cite{SchmidWR02} to verify Byzantine agreement
with link failures, in addition to process failures.
Isabelle/HOL~\cite{NipkowPW02} was used to certify the correctness of
several protocols encoded in the heard-of model in~\cite{Charron-BostS09,DebratM12}.

These semi-automated proofs require a great amount of human
intervention and understanding of the distributed algorithms.
IVy~\cite{PadonMPSS16,McMillanP20} is an interactive verification tool,
whose goal is achieving higher automation
when producing a formal proof of a distributed algorithm,
by reducing the amount of human guidance as much as possible.
The main idea of the IVy methodology is to encode the algorithms
in the effectively-propositional fragment (EPR) of first-order logic.
It is not always straightforward to encode distributed algorithms and
their verification conditions in EPR, but once it is done, the
verification condition check is fully automatic.

\subsection{Verification of Distributed Systems}%
\label{sec:verifdistsys}
IronFleet~\cite{HawblitzelHKLPR17} implements a variant of
MultiPaxos~\cite{Lamport98} in Dafny~\cite{Leino10}, which allows for Hoare-logic style
program verification.
Verdi~\cite{WilcoxWPTWEA15} verifies an implementation of the Raft protocol~\cite{OngaroO14}
using the Coq proof assistant~\cite{coq}, and translates the Coq proof
into a verified implementation in OCaml.
Chapar~\cite{LesaniBC16} also uses the Coq to OCaml translation to obtain a verified
implementation of a distributed key-value store.
When verifying implementations of distributed systems,
one has to be very careful about the assumptions,
about the calls to unverified external libraries,
and about the correctness of the specifications themselves.
An empirical study~\cite{FonsecaZWK17} analyzed the three verified
implementations produced by IronFleet, Verdi, and Chapar, and
reported existence of bugs in the interfaces between the verified code
and the unverified external libraries or operating system.

Recently, it was observed that in order to automatically verify asynchronous
distributed programs, one can define reductions~\cite{EF82,Chaouch-SaadCM09}
in order to reduce reasoning about an asynchronous system
to reasoning about a synchronous system, equivalent to the
original one~\cite{DDMW19:CAV}.
Due to the non-determinism that comes from the faults and the
asynchronous computation model,
these synchronized versions of the asynchronous programs
have different fault and computation semantics to those considered in this thesis.
We list some of the approaches based on the idea of reductions.
PSync~\cite{DragoiHZ16} was introduced as a domain-specific language for specifying
and implementing fault-tolerant distributed algorithms, which is based on
the heard-of model and  can be translated to the consensus logic of~\cite{DHVWZ14},
and thus the same invariant checking techniques can be applied.
A decision procedure for invariant checking of a given asynchronous message-passing
program, whose computations can be reduced to computations of an equivalent
round-based program with a bounded number of send operations per round,
is presented in~\cite{BouajjaniEJQ18}.
Several methods based on Lipton reduction~\cite{Lipton75} were introduced in
order to reduce asynchronous programs to other programs, for which reasoning is easier.
For example,~\cite{BakstGKJ17} introduces canonical sequentialization, which is
a sequential program, equivalent to a given asynchronous message-passing program,
and~\cite{KraglEHMQ20} defines inductive sequentialization, a sequential program
to which a given asynchronous program is reduced by combining reduction, abstraction,
and inductive reasoning.
Asynchronous programs are reduced to equivalent synchronous programs in~\cite{KraglQH18},
and their invariants checked using a dedicated model checker.
Another technique for reducing asynchronous to synchronous programs was given
in~\cite{GleissenthallKB19}, where the correctness of the obtained synchronous program is
established using Hoare-style verification conditions and SMT solvers.

\subsection{Extending Threshold Automata}%
\label{sec:relatedTA}

We introduced threshold automata as a concise representation and a
     precise semantics for fault-tolerant distributed algorithms.
The verification results presented in this thesis typically apply only
     to threshold automata of specific shape, e.g., without increment
     of shared variables in a rule that appears in a loop.
The restrictions that we imposed on the form of allowed threshold
     guards and  the corresponding actions are inspired by the typical
     patterns seen in the benchmarks.
That is, these restrictions still allow us to model a class
     fault-tolerant distributed algorithms.
From a theoretical viewpoint there remained the question, whether
     lifting one or the other restriction maintains or breaks
     decidability of the verification problem.

The work presented in~\cite{KKW18:concur} explores various relaxations
     of the standard  restrictions, such as non-linear threshold
     guards, guards that compare  shared variables, decrementing
     shared variables, or incrementing them  inside self-loops.
For each of these theoretical extensions, the authors investigate the
     existence of a bounded diameter and decidability of reachability
     properties.
For the standard setting, a systematic analysis of computational
     complexity  of verification and synthesis in threshold automata
     has been conducted  in~\cite{BEL:atva20}.
The authors express the reachability relation as a formula in
     existential  Presburger arithmetic, and therefore prove that
     coverability and  reachability problems, as well as model
     checking of~\ELTLTB{}  properties are NP-complete, while
     synthesizing threshold guards  is~$\Sigma^2_p$-complete.

\paragraph{Acknowledgments.} This survey is based on multiple results of
     a long-term
research agenda~\cite{JohnKSVW13:fmcad,KLVW17:FMSD,KLVW17:POPL,LKWB17:opodis,bertrand_et_al19,StoilkovskaKWZ19,SKWZ21:sttt}.
We are grateful to our past and present
collaborators Nathalie Bertrand, Roderick Bloem, Annu Gmeiner, Jure Kukovec, Ulrich Schmid,
Helmut Veith, and Florian Zuleger, who contributed to many of the described
ideas that are now implemented in ByMC\@.

\bibliographystyle{alphaurl}
\bibliography{lada,lit}

\clearpage

\appendix

\section{Abstraction of Tendermint Consensus Round in TLA+}\label{sec:TATLA}

\includegraphics[width=0.95\textwidth]{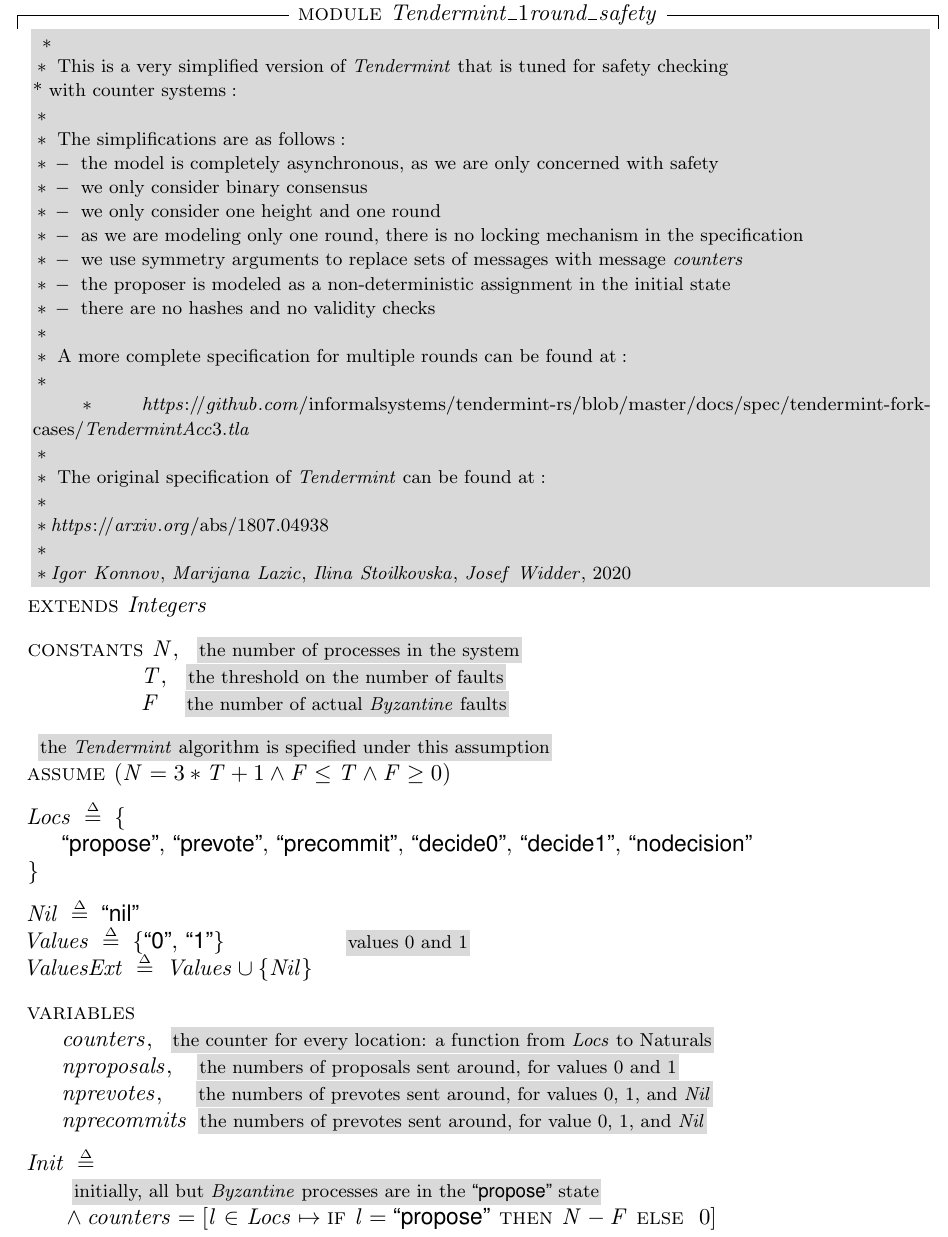}

\includegraphics[width=0.95\textwidth]{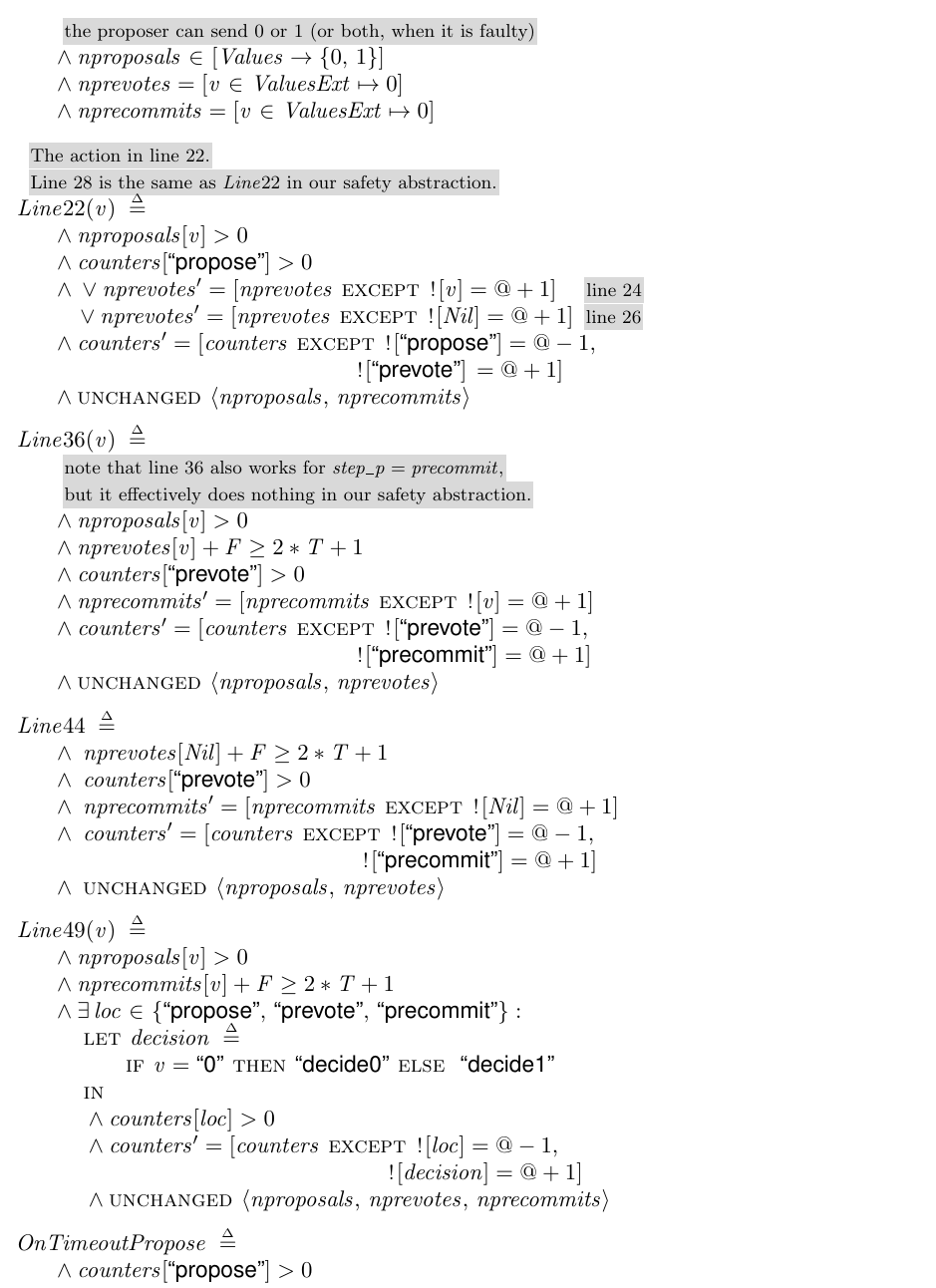}

\includegraphics[width=0.95\textwidth]{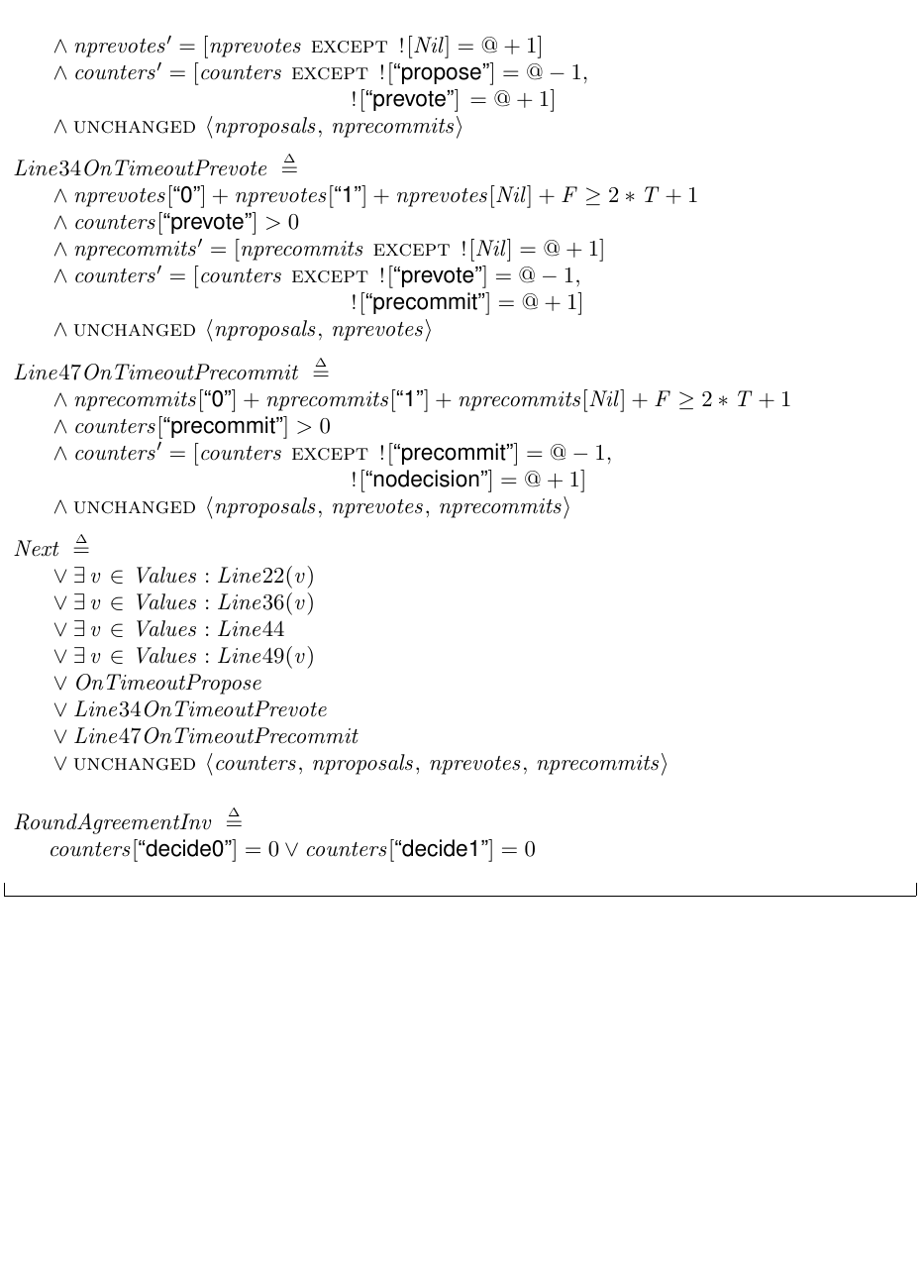}

\end{document}